\documentclass[a4paper,11pt]{article}
\usepackage{jheppub}
\usepackage{romannum}

\usepackage{booktabs}
\usepackage{siunitx}
\usepackage{subcaption}
\usepackage{mathtools}
\usepackage{xcolor}
\usepackage{fancyvrb}

\newcommand{\powhegbox}{\texttt{POWHEG\,BOX\,V2}}
\newcommand{\powheg}{POWHEG}
\newcommand{\hooks}{\texttt{PowhegHooks}}
\newcommand{\pythia}{\texttt{PYTHIA\,8}}
\newcommand{\sherpa}{\texttt{SHERPA}}
\newcommand{\hera}[1]{HERA~\Romannum{#1}}
\newcommand{\MSbar}{$\overline{\text{MS}}$}
\newcommand{\DISg}{$\text{DIS}_\gamma$}
\newcommand{\ie}{i.e.}
\newcommand*\diff{\mathop{}\!\mathrm{d}}
\newcommand{\bphi}{{\bf \Phi}}

\preprint{MS-TP-26-31}

\title{Dijet photoproduction in \boldmath$ep$ and \boldmath$eA$ collisions in \texttt{POWHEG}}

\author[a]{A. Feike,}
\author[a]{T.~Je\v{z}o,}
\author[a]{M. Klasen}

\affiliation[a]{Institut  für  Theoretische  Physik,  Universität  Münster,  Wilhelm-Klemm-Straße 9, 48149 Münster, Germany}

\emailAdd{alex.feike@uni-muenster.de}

\abstract{We present the implementation of dijet photoproduction in electron-proton and electron-nucleus collisions in the \powhegbox{} Monte Carlo event generator. The process is computed at next-to-leading order in QCD including both direct and resolved photon contributions and is consistently matched to parton showers using the \powheg{} method. The implementation is validated by using \powhegbox{} internal consistency checks as well as by comparing fixed-order results with \sherpa. We compare our predictions to dijet photoproduction data from DESY HERA I and II, finding good agreement within theoretical uncertainties, and provide predictions for both electron-proton and electron-gold collisions at the upcoming BNL EIC, illustrating its potential to further constrain both photon and nuclear parton distribution functions.}

\begin{document}
\maketitle

\section{Introduction}

At high energies, photons colliding with hadrons exhibit a rich internal structure. On the one hand, the pointlike component of the photon gives rise to a direct electromagnetic interaction with the partons of the hadronic target. On the other hand, the photon itself reveals a hadronic structure, originating from its perturbative splitting into quark-antiquark pairs and from fluctuations into light neutral vector mesons as described non-perturbatively by the vector-meson-dominance (VMD) model~\cite{Stodolsky:1966am,Sakurai:1972wk}. The direct and resolved components are theoretically closely related, since the collinear singularity of the pointlike contribution must be regularized, e.g.\ dimensionally, and then absorbed into the photon parton distribution functions (PDFs)~\cite{Klasen:2002xb}. As a consequence, the separation into direct and resolved contributions is dependent on the factorization scheme and scale, and only their sum constitutes a physically well-defined quantity.

In lepton-hadron and lepton-lepton collisions, the incoming lepton can radiate a photon. For small photon virtualities $Q^2$, this photon can be approximated to be quasi-real, defining the photoproduction regime, in contrast to deep-inelastic scattering (DIS) at large $Q^2$. The resulting flux of quasi-real photons can then be described as a function of the longitudinal photon momentum fraction and virtuality by the Weizsäcker-Williams (or equivalent-photon) approximation (WWA/EPA)~\cite{vonWeizsacker:1934nji,Williams:1934ad}. Measurements of the photon structure function $F_2^\gamma$ in inclusive DIS at $e^+e^-$ colliders provide information on the photon PDFs, albeit mostly on the (light) quarks and much less on the gluon distribution. Photon PDFs have been fitted to data by several groups, in particular GRV~\cite{Gluck:1991jc}, CJK~\cite{Cornet:2002iy,Cornet:2004nb}, SaS~\cite{Schuler:1995fk,Schuler:1996fc}, SAL~\cite{Slominski:2005bw} and Valo~\cite{Chithirasreemadam:2026mqp}.

Among other photoproduction channels, photoproduced jets have been most intensely studied both in $e^+e^-$ collisions at CERN's Large Electron-Positron collider (LEP)~\cite{DELPHI:2008fup} and in $ep$ collisions by the two experiments H1~\cite{H1:2006rre,H1:2002apm} and ZEUS~\cite{ZEUS:2001zoq,ZEUS:2012pcn,ZEUS:2007njl} at DESY's Hadron Electron Ring Accelerator (HERA). These measurements were accompanied on the theoretical side by several independent predictions at next-to-leading order (NLO) in QCD~\cite{Aurenche:1992sb,Aurenche:1993pk,Aurenche:1994mw,Aurenche:2000nc,Harris:1997hz,Harris:1997ji,Frixione:1997ks,Klasen:1995ab,Klasen:1995xe,Klasen:1996it,Klasen:1997br}. In addition to serving as a valuable tool for testing our understanding of the Standard Model (SM), they can provide additional constraints on the hadronic component of the photon, in particular the contribution of the gluon. Compared to the proton, however, the PDFs of the photon have remained considerably less constrained, owing to the limited amount of available data, the restricted kinematic coverage, and insufficient theoretical tools both in terms of precision and in terms of speed, leaving in particular the gluon distribution at low momentum fractions subject to sizeable uncertainties.

While the photon structure function measured in deep inelastic electron-photon scattering at $e^+e^-$ colliders constitutes one of the main observables used to fit these photon PDFs~\cite{Nisius:1999cv}, jet photoproduction provides complementary sensitivity, in particular to the gluon distribution of the photon. Dijet final states are well suited for these studies, since the two leading jets allow for a reconstruction of the underlying parton kinematics. In particular, the observed momentum fraction $x_{\gamma}^{\text{obs}}$ of the parton entering the hard scattering from the photon side (defined below) can be reconstructed from the jet observables. Since direct processes correspond to the full photon momentum entering the hard interaction ($x_{\gamma}^{\text{obs}} \rightarrow 1$) and resolved processes populate the region of smaller momentum fraction, this observable also provides an experimental handle to separate the two contributions~\cite{Butterworth:2005aq}.

With the advent of CERN's LHC and the end of the DESY HERA program, proton-proton collisions moved more into focus, while photoproduction processes receded into the background. However, the announcement of an Electron-Ion Collider (EIC)~\cite{AbdulKhalek:2022hcn} at BNL and discussions of a possible Future Circular Electron-Positron Collider (FCC-ee)~\cite{FCC:2018evy} at CERN have renewed interest in this field. This is reflected in new activity on jet photoproduction in state-of-the-art Monte Carlo generators like \texttt{PYTHIA} and \texttt{SHERPA}~\cite{Helenius:2017aqz,Hoeche:2023gme,Helenius:2024rth}. Compared to HERA, the EIC not only delivers higher luminosity, and thus improved statistics, but also extends the study from protons to heavy ions, providing a clean handle on nuclear effects~\cite{Klasen:2017kwb,Klasen:2018gtb,Guzey:2020zza}. Such measurements probe nuclear modifications of the parton distributions and are therefore of importance for the determination of nuclear PDFs, as provided by collaborations such as \texttt{nCTEQ} \cite{nCTEQ:2026ewr}, \texttt{EPPS} \cite{Eskola:2021nhw} and \texttt{nNNPDF} \cite{AbdulKhalek:2022fyi}.

An accurate theoretical description of such observables requires predictions that not only achieve precise fixed-order accuracy in perturbation theory, but also describe complete, physical final states down to the hadronic level, allowing a direct comparison with experimental data. This is achieved by matching NLO QCD calculations to parton showers, thus combining the hard matrix element calculations of the fixed-order result with the resummation of soft and collinear emissions and the subsequent hadronization provided by the shower while avoiding double-counting of radiation.

In this paper, we present the implementation of dijet photoproduction in the state-of-the art Monte Carlo event generator \powhegbox~\cite{Frixione:2007vw,Alioli:2010xd}. The process is computed at NLO in QCD and consistently matched to the parton shower using the \powheg{} method~\cite{Nason:2004rx}. This provides events for electron-proton and electron-nucleus scattering including both the direct and resolved photon contributions, offering a flexible tool for the interpretation of present and future photoproduction data.

The paper is organized as follows: In Sec.~\ref{sec: framework}, we briefly review the theoretical framework of factorization in photoproduction processes. Sec.~\ref{sec: implementation} describes the implementation of dijet photoproduction in the \powhegbox{}, including the minimal input for a new process, as well as modifications to the core code. We present numerical results in Sec.~\ref{sec: results}, covering fixed-order validation, shower variations, comparison to HERA data and predictions for the EIC. Finally, we summarize in Sec.~\ref{sec: conclusion}.

\section{Theoretical framework}
\label{sec: framework}

High-energy hadronic cross sections are computed within the framework of QCD factorization, which separates the process into a long-distance, non-perturbative part and a short-distance, perturbatively calculable one~\cite{Collins:1989gx}. For an electron-hadron collision with incoming momenta $P_1$ and $P_2$, respectively, this allows the cross-section to be formulated as a convolution of the parton distribution function (PDF) of the incoming hadron with an electron-parton hard-scattering cross-section,
\begin{equation}
    \sigma_{eH} = \sum\limits_j \int \diff x_{j/H} ~ f_j^{(H)}(x_{j/H},\mu_F) ~ \hat{\sigma}_{ej}(P_1,x_{j/H}P_2,\mu_F,\mu_R) ~,
\end{equation}
where $f_j^{(H)}$ is the PDF describing the probability of finding a parton $j$ carrying a longitudinal momentum fraction $x_{j/H}$ in the hadron $H$. Beyond leading order (LO), the partonic cross section develops collinear singularities associated with emissions from the incoming parton. Factorization allows these initial-state singularities to be absorbed into the PDFs, so that the hadronic cross section remains finite, at the price of introducing a dependence on an unphysical factorization scale $\mu_F$. Similarly, the renormalization of ultraviolet divergences introduces a dependence on the unphysical renormalization scale $\mu_R$. The PDFs are universal, such that the same distributions extracted in a set of processes in a given kinematic region can be used to predict other processes in different kinematic regions.

Since the electron is a fundamental particle, the cross section $\hat{\sigma}_{ei}$ is in principle calculable in perturbation theory. In the limit of small electron deflection angles, however, the Weizsäcker-Williams approximation (WWA), also known as equivalent-photon approximation (EPA), offers a convenient description. Its underlying idea is that the electromagnetic field of the electron, described in QED by an exchanged virtual photon, can be approximated at small scattering angles, and thus small virtualities, by a flux of \emph{quasi-real} photons that subsequently enter the hard process. In this limit, the electron-hadron cross-section factorizes further into
\begin{equation}
    \sigma_{eH}=  \sum\limits_j \int \diff{x_{\gamma/e}} \diff x_{j/H} ~ f_{\gamma}^{(e)}(x_{\gamma/e}) ~ f_j^{(H)}(x_{j/H},\mu_F) ~ \hat{\sigma}_{\gamma j}(x_{\gamma/e}P_1,x_{j/H}P_2,\mu_F,\mu_R)~,
\end{equation}
where $\hat{\sigma}_{\gamma j}$ is the cross-section for the interaction of a real photon with the parton in the target hadron and $x_{\gamma/e}$ denotes the longitudinal momentum fraction of the electron carried by the radiated quasi-real photon. Although this factorization looks formally similar to the hadronic case, the origin of the photon flux is quite different. Rather than being extracted from data with a non-perturbative framework, as is the case for PDFs, the photon flux follows from a perturbative calculation supplemented by the small electron scattering angle approximation and consequently does not involve a factorization scale. The approximation dates back to the independent works of Weizsäcker~\cite{vonWeizsacker:1934nji} and Williams~\cite{Williams:1934ad}. It was later refined into the improved WWA introduced in Ref.~\cite{Frixione:1993yw}, which supplements the leading logarithmic term by a contribution proportional to the electron mass. The resulting photon flux reads\footnote{Since the four-momentum $q$ of the virtual photon is spacelike ($q^2<0$), we adopt the conventional positive virtuality $Q^2=-q^2$. This interchanges the roles of the integration limits, ${Q^2_{\text{max}}= -q^2_{\text{min}} }$ and ${Q^2_{\text{min}}= -q^2_{\text{max}} }$.}
\begin{align}
    &f_\gamma^{(e)}(x_{\gamma/e})= \frac{\alpha_{\text{em}}}{2 \pi} \left[ \frac{1+\left(1-x_{\gamma/e} \right)^2}{x_{\gamma/e}} \ln \frac{Q^2_{\text{max}}}{Q^2_{\text{min}}}+ 2m_e^2 x_{\gamma/e} \left(\frac{1}{Q^2_{\text{max}}} - \frac{1}{Q^2_{\text{min}}}\right) \right]~. 
\end{align}
The minimal photon virtuality is fixed by kinematics,
\begin{equation}
    Q^2_{\text{min}}=\frac{\left(m_e x_{\gamma/e}\right)^2}{1-x_{\gamma/e}} ~,
\end{equation}
whereas the maximal virtuality is not fixed by the process itself, but is characterized by the experimental setup. It is commonly specified either directly through a fixed value or in terms of the maximal deflection angle $\theta_c$, up to which the exchanged photon is still regarded as quasi-real. Our implementation supports both options and, similarly to \sherpa~\cite{Sherpa:2024mfk,Hoeche:2023gme}, combines them into a single upper bound, if both are given by the user:
\begin{equation}
    Q^2_{\text{max}}=\text{max} \left(Q^2_{\text{min}}+E^2(1-x_{\gamma/e}) \theta_c^2,~ Q^2_{\text{max,user}}\right)~.
\end{equation}
Treating the quasi-real photon as a structureless particle, however, captures only part of its dynamics. Beyond LO, the partonic cross-section $\hat{\sigma}_{\gamma j}$ develops collinear singularities associated with the photon splitting into a $q\bar{q}$ pair. As in the hadronic case, these singularities are absorbed into a photon PDF $f_i^{(\gamma)}(x_{i/\gamma},\mu_F)$, describing the probability of finding a parton $i$ with longitudinal momentum fraction $x_{i/\gamma}$ in the real photon. Photon interactions thus separate into a direct and a resolved component, formally distinguished by yet another factorization step,
\begin{align}
    \begin{split}
        \sigma^{(\text{direct})}_{eH}
        &= \sum_j \int \diff{x_{\gamma/e}} \diff x_{j/H}
        ~ f_{\gamma}^{(e)}(x_{\gamma/e}) ~ f_j^{(H)}(x_{j/H},\mu_F) \\
        &\qquad \times \hat{\sigma}_{\gamma j}\left(x_{\gamma/e}P_1,x_{j/H}P_2,\mu_F,\mu_R\right),
    \end{split}
    \label{eq. factorisation direct}\\
    \begin{split}
        \sigma^{(\text{resolved})}_{eH}
        &= \sum_j \int \diff{x_{i/\gamma}} \diff{x_{\gamma/e}} \diff x_{j/H}
        ~ f_i^{(\gamma)}(x_{i/\gamma},\mu_F) ~ f_{\gamma}^{(e)}(x_{\gamma/e}) \\
        &\qquad \times f_j^{(H)}(x_{j/H},\mu_F)
        ~\hat{\sigma}_{ij}\left(x_{i/\gamma}x_{\gamma/e}P_1,x_{j/H}P_2,\mu_F,\mu_R\right).
    \end{split}
    \label{eq. factorisation resolved}
\end{align}
For brevity, we assigned the hadron and the photon PDF the same factorization scale, which is a choice. The scale evolution obeys an inhomogeneous DGLAP equation, an additional term with respect to the hadronic case arising from the ${\gamma\rightarrow q \bar{q}}$ splitting. The solution combines a perturbatively generated point-like component with a non-perturbative hadronic component, the latter commonly modeled via vector-meson dominance.

Most available photon PDF sets are formulated in the \DISg{} factorization scheme, which absorbs the point-like splitting into the quark distributions contributing to the photon structure function $F_2^\gamma$ ~\cite{Gluck:1991ee}. Within \powhegbox, however, all matrix elements are computed in the \MSbar{} scheme. We therefore apply the corresponding \DISg$\to$\MSbar{} conversion~\cite{Gluck:1991jc} by default to every \DISg{} set in both LO and NLO runs. This behavior can be disabled by the user through a dedicated switch.

An overview of the implemented photon PDF sets, together with the native factorization schemes and implemented ranges on momentum fraction and factorization scale, is given in Tab.~\ref{tab: photonPdf}. The \texttt{VALO} PDF set is listed for completeness, although, since it is provided in the LHAPDF format, no dedicated implementation was required. Consequently, any future photon PDF set published via the LHAPDF library can be used out of the box. This applies, e.g., to the \texttt{PL$\gamma$26} set~\cite{Buckley:2026tbf}, which appeared shortly before this work was completed and is therefore not considered in any PDF comparisons in the following sections.

\begin{figure}
    \centering
    \begin{subfigure}{0.39 \textwidth}
        \centering
        \includegraphics[scale=0.7]{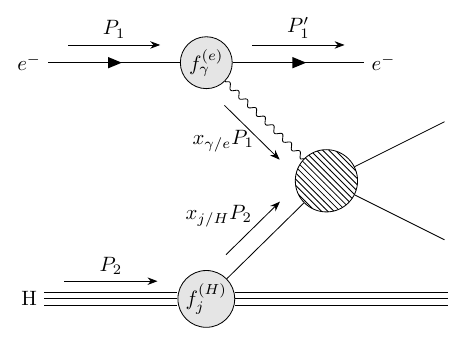}
        \caption{direct}
    \end{subfigure}
    \hfill
    \begin{subfigure}{0.6 \textwidth}
        \centering
        \includegraphics[scale=0.7]{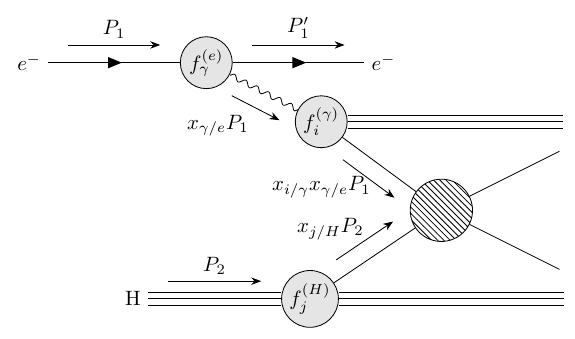}
        \caption{resolved}
    \end{subfigure}
    \caption{Photoproduction diagrams illustrating ${2\to2}$ parton scattering in $eH$ collisions after factorization, including the relevant fluxes and momentum fractions, for direct (a) and resolved (b) contributions. Remnant jets from the incoming photon and hadron $H$ are shown explicitly.}
    \label{fig:photoproductionSketch}
\end{figure}

The implemented processes are shown diagrammatically in Fig.~\ref{fig:photoproductionSketch}. It illustrates the incoming lepton and hadron beams, the resolution of the lepton into a quasi-real photon, and, in the resolved case, the further resolution into its partonic substructure. Analogously, the hadron is resolved into a parton via the proton (or nuclear) PDFs.

\begin{table}
    \centering
    \begin{tabular}{l l l l l }
        \toprule
        PDF & PDF set  & Scheme & $x$-range & $\mu_F^2$-range in \si{\giga \electronvolt ^2}   \\
        \midrule
        GRV~\cite{Gluck:1991jc}        
            & \texttt{GRV\,LO}    & -                     & $[\SI{e-5}{},1]$  & $[0.25,\SI{e6}{}]$\\
            & \texttt{GRV\,HO}    & \DISg   & $[\SI{e-5}{},1]$  & $[0.3, \SI{e6}{}]$\\
        CJK~\cite{Cornet:2002iy,Cornet:2004nb} 
            & \texttt{CJKL}                          & -                     & $[\SI{e-5}{},1]$  & $[1,\SI{2e5}{}]$\\
            & \texttt{CJK\,NLO}                       & \DISg   & $[\SI{e-5}{},1]$  & $[1,\SI{2e5}{}]$\\
            & $\text{\texttt{FFNS1}}_\text{\texttt{CJK}}$     & \DISg   & $[\SI{e-5}{},1]$ & $[1,\SI{2e5}{}]$\\
            & $\text{\texttt{FFNS2}}_\text{\texttt{CJK}}$     & \DISg   & $[\SI{e-5}{},1]$ & $[1,\SI{2e5}{}]$\\
        SAL~\cite{Slominski:2005bw} 
            & \texttt{SAL}   & \DISg   & $[\SI{e-5}{},0.9999]$ & $[1.8,\SI{e5}{}]$\\
        SaS~\cite{Schuler:1995fk,Schuler:1996fc}
            & \texttt{SAS1M}     & \MSbar    & $[\SI{e-4}{},1]$ & $[0.36,\SI{1.44e6}{}]$\\
            & \texttt{SAS2M}     & \MSbar    & $[\SI{e-4}{},1]$ & $[4,\SI{16e6}{}]$\\
            & \texttt{SAS1D}     & \DISg       & $[\SI{e-4}{},1]$ & $[0.36,\SI{1.44e6}{}]$\\
            & \texttt{SAS2D}     & \DISg       & $[\SI{e-4}{},1]$ & $[4,\SI{16e6}{}]$\\
        VALO1.0~\cite{Chithirasreemadam:2026mqp}
            & \texttt{VALO10\_LO}            & -                         & $[\SI{1e-5}{1},]$ & $[1,\SI{1e5}{}]$\\
            & \texttt{VALO10\_NLO\_MSbar}    & \MSbar    & $[\SI{1e-5}{1},]$ & $[1,\SI{1e5}{}]$\\
            & \texttt{VALO10\_NLO\_DISg}      & \DISg       & $[\SI{1e-5}{1},]$ & $[1,\SI{1e5}{}]$\\
        \bottomrule
    \end{tabular}
    \caption{Overview of the photon PDFs currently implemented in our code. For each PDF set, we list the factorization scheme and the supported ranges in $x$ and $\mu_F^2$. Entries with no scheme specified correspond to LO fits.}
    \label{tab: photonPdf}
\end{table}

\section{Implementation in \powhegbox}
\label{sec: implementation}

The Monte Carlo generator provided by the \powhegbox{} framework~\cite{Frixione:2007vw,Alioli:2010xd} produces unweighted events at NLO precision following the \powheg{} method~\cite{Nason:2004rx}. To this end, the \powhegbox{} decomposes the real-emission contribution into a sum of terms,
\begin{equation}
    R =\sum_{\alpha_r} R^{\alpha_r} \, , \label{eq. reals separated into singular regions}
\end{equation}
each of which is singular in one, and only one, singular region $\alpha_r$ of the $(n+1)$-particle phase space $\bphi_{n+1}$. Every singular region is associated with an $n$-particle \emph{underlying Born} flavor structure $f_b$, obtained by merging the soft parton with its emitter or the two collinear partons into one single parton. The matching of the flavor structures is supplemented by mapping the relevant phase spaces, \ie~of the real-emission phase space $\bphi_{n+1}$ in the singular region to an underlying Born phase space $\bar{\bphi}^{\alpha_r}_{n}$ and a factorized radiation phase space $\Phi^{\alpha_r}_{\text{rad}}$:
\begin{equation}
    \bphi_{n+1} \longleftrightarrow \bar{\bphi}^{\alpha_r}_{n} \Phi^{\alpha_r}_{\text{rad}} ~.
\end{equation}
The central idea of the \powheg{} method is to generate an underlying Born configuration first and to subsequently generate the hardest emission from it. To achieve NLO accuracy, the underlying Born configurations are sampled according to the NLO inclusive $n$-particle cross section, given as
\begin{equation}
    \begin{split}
        \bar{B}^{f_b}(\bphi_n) = &
        \left[B(\bphi_n) + V(\bphi_n) \right]
        + \sum \limits_{\alpha_r \in \{\alpha_r|f_b\}} \int \left [
        \diff \Phi_{\text{rad}} \left\{R(\bphi_{n+1}) - C(\bphi_{n+1}) \right\}
        \right]_{\alpha_r}^{{\bf \bar{ \Phi}}_n^{\alpha_r}=\bphi_n}
        \\
        & +  \sum \limits_{\alpha_{\oplus} \in \{\alpha_\oplus|f_b\}} \int \frac{\diff z}{z} G^{\alpha_\oplus}_\oplus (\bphi_{n,\oplus})
        +  \sum \limits_{\alpha_{\ominus} \in \{\alpha_\ominus|f_b\}} \int \frac{\diff z}{z} G^{\alpha_\ominus}_\ominus (\bphi_{n,\ominus}) ~,
    \end{split}
    \label{eq. Bbar}
\end{equation}
in which all radiative degrees of freedom have been integrated out. Here, $B$ denotes the Born contribution, $V$ the finite part of the virtual corrections, $C$ the real counterterms, and $G_\oplus$ and $G_\ominus$ the initial-state collinear remnants associated with the incoming legs. The notation ${[\, \cdots \, ]_{\alpha_r}^{{\bf \bar{ \Phi}}_n^{\alpha_r}=\bphi_n}}$ indicates that the expression in the brackets is evaluated in the singular region $\alpha_r$, with the underlying Born kinematics fixed to $\bphi_n$.

The generation of the hardest emission is then described by the \powheg{} master formula
\begin{equation}
    \begin{split}
        d \sigma = &
        \sum_{f_b} \bar{B}^{f_b}(\bphi_n) \diff \bphi_n \Bigg \{ \Delta ^{f_b}(\bphi_n,p_T^{\text{min}})
        \\
         & + \sum \limits_{\alpha_r \in \{\alpha_r|f_b\}}
          \frac{\left[ \diff \Phi_{\text{rad}} \theta(k_T(\bphi_{n+1})-p_T^{\text{min}}) \Delta^{f_b}(\bphi_n,k_T) R(\bphi_{n+1}) \right]_{\alpha_r}^{{\bf \bar{ \Phi}}_n^{\alpha_r}=\bphi_n}}{B^{f_b}(\bphi_n)}
        \Bigg\} ~, \label{eq. pwhg master formula}
    \end{split}
\end{equation}
where the Sudakov form factor is given by
\begin{equation}
    \Delta^{f_b}(\bphi_n, p_T) 
    =\exp \left\{ - \sum \limits_{\alpha_r \in \{\alpha_r|f_b\}} \int
    \frac{\left[ \diff \Phi_{\text{rad}} R(\bphi_{n+1}) \theta(k_T(\bphi_{n+1})-p_T) \right]_{\alpha_r}^{{\bf \bar{ \Phi}}_n^{\alpha_r}=\bphi_n}}{B^{f_b}(\bphi_n)}
    \right\} ~. \label{eq: modified sudakov powheg}
\end{equation}
The first term in the curly brackets of Eq.~(\ref{eq. pwhg master formula}) corresponds to events without any resolved emission. Here, the Sudakov form factor encodes the probability that no emission with transverse momentum above the cutoff scale $p_T^{\text{min}}$ occurs. The second term describes the generation of the hardest emission with transverse momentum ${k_T>p_T^{\text{min}}}$. It's emission probability follows $R/B$, while the Sudakov factor ensures that no harder particle is emitted. Since the curly bracket integrates to unity, the \powheg{} procedure preserves the NLO normalization given by $\bar{B}^{f_b}$.

Any subsequent shower, that supports $p_T$-ordered or -vetoed showers, may then be used to generate further emissions at scales lower than the \powheg{} emission scale $k_T$. To simulate a given process, the \powhegbox{} relies on a process implementation that supplies the corresponding process-specific information. Such implementations already exist for a wide range of processes, and new ones can be added by supplementing implementation for a set of required routines. A detailed description of all required ingredients can be found in Ref.~\cite{Alioli:2010xd}. Here, we do not repeat this description in full. Instead, we only briefly state how the required amplitudes are obtained and then discuss our implementation of the phase space generation suitable for both direct and resolved processes. Finally, we describe the modifications that had to be made to the \powhegbox{} itself.

\subsection{Amplitudes and phase space}
\label{sec: Amplitudes and phase space}

At the Born level, we implement the flavor structures ${\gamma j \rightarrow jj}$ and ${jj\rightarrow jj}$ for the direct and resolved contribution, respectively, where $j$ is a placeholder for a gluon or any massless quark. Since we work in the five-flavor number scheme, this comprises all quarks but the top. For the real corrections, we add one final-state parton, \ie~we implement ${\gamma j \rightarrow jjj}$ and ${jj\rightarrow jjj}$.

For each contributing partonic subprocess, the \powhegbox{} requires on the amplitude side the Born squared matrix element together with its color-correlated and spin-correlated counterparts, the real-emission squared matrix elements, and the finite part of the one-loop virtual corrections. For the resolved contributions, all of these partonic amplitudes coincide with those of the already available \texttt{dijet} process~\cite{Alioli:2010xa}. Thus, we simply link to the existing routines whenever the partonic process contains no photon. The direct contribution, in contrast, was recalculated with \texttt{FORMCALC}~\cite{Hahn:1998yk} and implemented in new routines. As in the \texttt{dijet} process, the Born matrix elements already exhibit infrared (IR) singularities, which can be treated either with a Born suppression factor or a generation cut \texttt{bornktmin}, the latter of which we use throughout this work.

To validate the calculated amplitudes, we compared them with \texttt{MadGraph5\_aMC\@NLO}~\cite{Alwall:2014hca,Frederix:2018nkq} numerically. Additionally, we exploited the fact that the direct subprocesses are related by crossing to those of prompt-photon production. Crossing the photon into the final state, we reproduced the corresponding Born, real and virtual matrix elements of the \texttt{directphoton} implementation in \powhegbox~\cite{Jezo:2016ypn,Klasen:2017dsy}.
In both cases, the comparison of 3000 random phase space points showed excellent agreement to at least 5 digits and no systematic phase space dependence was detected.

For the implementation of the Born phase space, we use
\begin{equation}
    \diff x_A \diff x_B \diff \text{PS}_2=\frac{p_T}{4 \pi S} \diff p_T \diff y_1 \diff y_2 \frac{\diff \phi}{2 \pi}\eqqcolon \diff \Phi^{(1,1)} ~,\label{eq. dPS_11}
\end{equation}
where $x_A$ and $x_B$ denote the momentum fractions of the incoming partons relative to their parent beams $A$ and $B$, $p_T$ is the transverse momentum of the outgoing partons (identical for both in a ${2\to 2}$-process), $y_1$ and $y_2$ are their rapidities, $\phi$ is the azimuthal angle of the first outgoing parton, and ${S=4 E_AE_B}$ is the squared center-of-mass energy of the colliding beams. Momentum conservation in the transverse plane forces the outgoing particles to be back-to-back, \ie~the azimuthal angle of the second parton is fixed to ${\phi + \pi}$. Owing to the independence of the calculation of the azimuthal angle, the integral is carried out analytically and $\phi$ is reconstructed as a random value distributed uniformly in $[0,2\pi)$. 

The superscript $(1,1)$ indicates that each beam is resolved exactly once in the sense of the phase space parametrization, \ie~a single momentum fraction connects each beam to the parton entering the hard process. This picture is complete in case of incoming hadrons. 
In photoproduction, however, the electron beam can involve an additional resolution step, corresponding to the conventional direct/resolved distinction. In the direct case, the photon itself enters the hard process and $x_A=x_{i/\gamma}$. In the resolved case, however, the photon emitted by the electron is in turn resolved, so that the momentum fraction splits into
\begin{equation}
    x_A =x_{i/\gamma}^{(A)} \, x_{\gamma/e}^{(A)} \, . \label{eq:x-resolved}
\end{equation}
Using ${\diff x_{i/\gamma}^{(A)} \, \diff x_{\gamma/e}^{(A)} = \diff x_A \, \diff x_{\gamma/e}^{(A)} \, 1/x_{\gamma/e}^{(A)}}$ yields
\begin{equation}
    \diff \Phi^{(2,1)} = \diff x_{i/\gamma}^{(A)} \, \diff x_{\gamma/e}^{(A)} \, \diff x_B \diff \text{PS}_2= \frac{ \diff x_{\gamma/e}^{(A)} }{x_{\gamma/e}^{(A)}} \diff \Phi^{(1,1)} ~. \label{eq. dPS_21}
\end{equation}
Analogously, ${\diff \Phi^{(1,2)}}$ is obtained by resolving $B$ into parton $j$, while  ${\diff \Phi^{(2,2)}}$ contains both factors.

To cover all of these cases within a single implementation, we first sample $p_T$, $y_1$ and $y_2$ in the hadronic center-of-mass frame. The momentum fractions of the incoming partons then read
\begin{align}
    x_{A/B}=\frac{p_T}{\sqrt{S}}\sum\limits_{i=1}^2 \exp(\pm y_i)~,
\end{align}
yielding the momenta
\begin{align}
    p_{1/2}^{\mu}&=\frac{\sqrt{S}}{2}(x_{A/B},0,0,\pm x_{A/B}) 
    \quad \text{and} \\
    p_{3/4}^{\mu}&=p_T (\cosh{y_{1/2}},\pm \cos{\phi},\pm \sin{\phi},\sinh{y_{1/2}}) ~.
\end{align}
These momenta are subsequently boosted into the partonic center-of-mass frame and the laboratory frame.

At this stage, only the momenta of the beams and of the partons entering the hard process are known, \ie~the kinematics is agnostic to whether there is an intermediate resolved photon. Since the amplitudes are evaluated at the partonic level, this is sufficient to compute the partonic cross section. For the convolution with the fluxes, however, we have to distinguish whether the incoming parton is the photon itself or a parton resolved from the photon. Whenever \powhegbox{} requests the PDF of a lepton beam at momentum fraction $x_A$, the call is redirected to a newly implemented routine setting the corresponding flux. The photon-in-electron flux is simply the WWA flux, \ie~$f^{(e)}_{\gamma}(x_A)$, and no further sampling is required. For a parton, the lepton PDF is the convolution of the WWA flux and the photon PDF,
\begin{align}
    f^{(e)}_{i}(x_A, \mu_F) = \int\limits_0^1 \diff x_{\gamma/e} \, \int\limits_0^1\diff x_{i/\gamma}  \, f^{(e)}_{\gamma}(x_{\gamma/e}) \, f^{(\gamma)}_{i}(x_{i/\gamma}, \mu_F) \, \delta(x_A-x_{i/\gamma}x_{\gamma/e}) ~. \label{eq. f convolution}
\end{align}
Instead of evaluating the additional integral at every call, we treat $x_{\gamma/e}$ as an additional integration variable of the Monte Carlo. For a given $x_A$, a dedicated subroutine samples $x_{\gamma/e}$ within the range allowed by the photon PDF, fixes $x_{i/\gamma}$ following Eq.~\eqref{eq:x-resolved}, and returns the additional Jacobian of Eq.~\eqref{eq. dPS_21}. To account for the modified phase space of the resolved photon contribution, the product in Eq.~\eqref{eq. f convolution} is further multiplied by this additional Jacobian. The additional phase space dimension is thus entirely absorbed into the parton-in-lepton flux, such that direct and resolved contributions are generated with the same Born phase space subroutine.

\powhegbox{} constructs the real-emission kinematics from the underlying Born configuration and the three Frixione, Kunszt and Signer (FKS)~\cite{Frixione:1995ms,Frixione:1997np} radiation variables. For initial-state radiation (ISR), this changes the incoming momentum fractions. On the lepton beam side with a resolved photon, this seems to call for a modification. Physically, the emission takes place inside the photon, so one would keep $x_{\gamma/e}$ fixed and change only $x_{i/\gamma}$, which would require modifications of \powhegbox. This is, however, not necessary. The partonic matrix elements depend only on the total momentum fraction $x_A$, while its decomposition into $x_{\gamma/e}$ and $x_{i/\gamma}$ enters solely through the parton-in-electron flux, which is a convolution over all possible decompositions. As this convolution is evaluated by the Monte Carlo sampling in any case, it is irrelevant whether the decomposition for the real-emission kinematics is inherited from the Born configuration or sampled anew. 

Although this modification is not required, we nevertheless implemented it in order to validate the above argument numerically. To this end we modified \powhegbox{} such that the real emission kinematics are constructed from $x_{i/\gamma}$ alone, while $x_{\gamma/e}$ is inherited from the Born configuration and compared this setup to the unmodified one. At fixed-order, both setups give total cross sections that agree within the statistical precision of about \SI{0.05}{\percent} and the differential jet distributions agree bin by bin within their statistical uncertainties. We therefore stick with the unmodified phase space generation for the real emission of \powhegbox{} and no dedicated treatment for the resolved photon is required for the real emission. Whenever the flux is requested at the real-emission momentum fraction $x_A$, the same routine as described above for the Born configuration is called. It samples $x_{\gamma/e}$ and fixes $x_{i/\gamma}$ via Eq.~\eqref{eq:x-resolved}.

\subsection{Classification of singular regions}
\label{sec: Modifications to powhegbox}

In principle, direct and resolved contributions could be generated together in a single Monte Carlo run. In practice, however, it is simpler to treat them separately, as has traditionally been done. We follow this approach, keeping in mind that only the sum of both contributions is physically meaningful. Moreover, while the separation is unambiguous at the Born level, it is no longer clear-cut at NLO because of the splitting ${\gamma\rightarrow q \bar{q}}$. A prescription is therefore needed to assign the NLO contributions to either class. Since the underlying Born configuration $f_b$ plays a central role in \powhegbox{}, we classify each singular region $\alpha_r$ individually, rather than the full real flavor structure, according to its underlying Born process. A singular region is labeled resolved if its underlying Born process is resolved and direct if its underlying Born process is direct. While each singular region is defined by its singularity, it still carries of course also finite contributions.

\begin{figure}
    \centering
    \includegraphics[width=\textwidth]{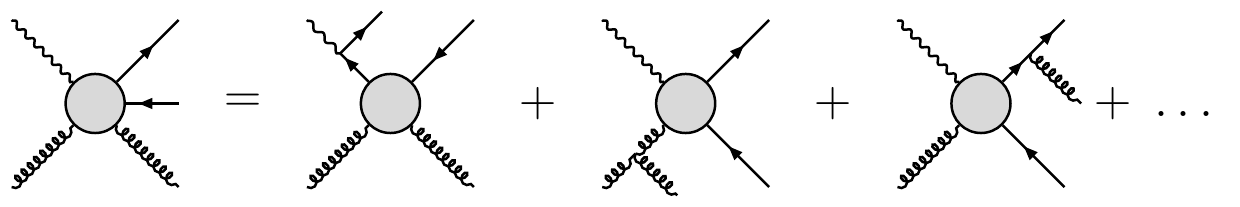}
    \caption{Schematic sketch of the division of a ${2\to3}$ process into singular regions.}
    \label{fig:singularRegions}
\end{figure}

In Fig.~\ref{fig:singularRegions} we schematically show the singular regions of the real-emission process ${\gamma g \rightarrow q \bar{q} g}$, illustrating the concept of Eq.~\eqref{eq. reals separated into singular regions}. Each diagram on the right-hand side corresponds to one such region, with the singular splitting drawn explicitly. If the outgoing quark becomes collinear to the incoming photon, the photon effectively splits into a ${q \bar{q}}$ pair. Thus the underlying Born process is ${\bar{q}g\rightarrow \bar{q}g}$ and this region is assigned to the resolved contribution. Regions defined by the gluon becoming soft or collinear to the incoming gluon or the outgoing (anti)quark build upon the Born ${\gamma g\rightarrow q \bar{q}}$ and are therefore considered direct. To perform this classification consistently, our implementation always constructs the complete set of Born and real-emission flavor structures, regardless of whether direct or resolved events are requested. As a consequence, the same set of singular regions are identified in both runs, and the decomposition of the real contribution in Eq.~\eqref{eq. reals separated into singular regions} contains the same terms in both cases. To make sure each run only evaluates the matrix elements in the regions whose underlying Born process matches the selected contribution, we introduce the logical functions \texttt{use\_alr} and \texttt{use\_born}. If they return \texttt{true}, the corresponding matrix element is calculated, else it is skipped. These checks are applied wherever the code loops over singular regions, as for example, in the sums entering the calculations of the $\bar{B}$ function, Eq.~\eqref{eq. Bbar}, the POWHEG master formula, Eq.~\eqref{eq. pwhg master formula}, or the Sudakov form factor, Eq.~\eqref{eq: modified sudakov powheg}.

Skipping singular regions is possibly not the only way to distribute the direct and resolved contributions over two separate runs. In Ref.~\cite{FerrarioRavasio:2020guj}, individual singular regions are suppressed by a damping factor. 
Our proposal is, however, particularly economical, since most ingredients required for the separation are already available within the \powhegbox{}.
We validated this method further by first explicitly verifying that the code identifies all singular regions associated with resolved production in the resolved run and all those associated with direct production in the direct run and that together they form the complete set of singular regions. Second, we used the internal check of the singular limits provided by the \powhegbox{}, which tests that the real emission matrix elements are correctly approximated by their soft (collinear) counterparts in the soft (collinear) limit. Although this check is a standard validation step for any process implementation, it is particularly useful here. For example, falsely skipping an underlying Born process that belongs to a non-skipped singular region results in a warning, revealing an incorrect division. Consistently, with the skipping method we obtain smooth integration grids, providing an additional indication that all singularities are correctly identified, classified, and subtracted. Third, we compared fixed-order predictions for three-jet observables, that receive no contribution from Born and virtual matrix elements, for the resolved and direct contribution obtained with our proposed splitting, including the skipping of regions, against a reference setup. In this setup, all Born flavor structures are enabled and only the real-emission flavor structures with an incoming photon or only those without one are included. The sum of the two reference runs agrees with the sum of our direct and resolved runs. This shows that also the non-singular parts associated with a singular region are distributed correctly. Together these checks confirm that the selection via skipping regions is mutually exclusive, the direct and resolved runs together reproduce the full NLO result without double counting. To our knowledge, this is the first time that selective skipping of singular regions was used to separate a sample into two separate contributions within \powhegbox.

As a result of the proposed separation into direct and resolved contributions, for a direct underlying Born process ${\gamma j \rightarrow jj}$ all associated radiations and therefore all singular regions originate from the emission of a QCD parton. On the other hand, the resolved process ${jj \rightarrow jj}$ may radiate QCD emission as well as QED emission via the initial-state splitting ${\gamma \rightarrow q \bar{q}}$. The event generation in \powhegbox{} is oblivious to whether an emission is of QED or QCD type. The modified Sudakov form factor (see Eq.~\eqref{eq: modified sudakov powheg}) is built from all singular regions $\alpha_r$ that have the same underlying Born flavor structure $f_b$. The hardest emission is then generated with the highest-bid method. In this case, effectively, an emission is generated for each $\alpha_r$ region, with an associated scale closely related to its $p_T$, but only the one with the largest scale is retained. Thus, for our resolved implementation, QED and QCD emissions compete, since the set of $\alpha_r$ regions contains both types. We stress that this competition of QED and QCD radiation is not an additional ingredient of our implementation, but arises by construction in \powhegbox{}.

However, this competition has direct consequences for the starting scale of the subsequent shower. While the emission is generated, trial emissions are drawn and accepted or rejected using a hit-or-miss approach, with the starting scale lowered for each new trial. The scale of the selected emission is then saved in the event file as the \texttt{SCALUP} value and marks the starting scale for the shower. This means that if the \powhegbox{} generates a QED emission on top of a resolved Born at scale \texttt{SCALUP}, no QCD emission has been accepted at a higher scale and vice versa. Hence, for the resolved underlying Born processes, QCD and QED emission scales are evolved simultaneously and \texttt{SCALUP} limits the hardness of both. For the direct underlying Born process, in contrast, only the QCD scale is evolved, so the \texttt{SCALUP} only limits the hardness of subsequent QCD radiation. How a QED shower should be added in this situation is handled differently in the literature. Ref.~\cite{Barze:2014zba} discusses the analogous case with QCD and QED interchanged and argues that neither choice of starting scale is fully satisfactory. Starting the shower at the scale of the underlying Born process allows emissions harder than the \powheg{} emission, potentially compromising the kinematic accuracy of the latter, which was generated assuming it to be the hardest emission. If it were instead started at \texttt{SCALUP}, QED emission harder than the \powheg{} emission would be missing entirely. Other studies, that consider both QCD and EW corrections~\cite{Granata:2017iod,Chiesa:2020ttl}, use NLO QCD events showered with a QCD+QED shower as a reference, where only the QCD shower is restricted by \texttt{SCALUP}, corresponding to the first of the two options above. In view of these different treatments, we follow a conservative approach and keep the QED shower to the minimum required to generate the $\gamma \to q \bar{q}$ splitting in the initial-state shower. The corresponding settings are given in Sec.~\ref{sec: Impact of the parton shower}.

An improved treatment would require generating QED emissions off the direct underlying Born process at the \powhegbox{} level, which is beyond the scope of the present work. As for the resolved contributions, \texttt{SCALUP} would then limit the hardness of both QCD and QED radiation and could thus serve as the common starting scale for both the QCD and QED parton shower. Note, however, that QED emissions off the direct process are suppressed by $\alpha/\alpha_S$ relative to the QCD emissions. Moreover, they are of minor importance, since the only relevant splitting ${\gamma \rightarrow q \bar{q}}$ is absent by construction in the direct contribution. The described competition treatment is closely related to the \emph{competition} (C) scheme introduced in Ref.~\cite{Barze:2014zba} for $W\gamma$ hadroproduction. There this scheme is compared to a \emph{no competition} (NC) approach, in which events with a $Wj$ underlying Born process radiates only a photon at the \powheg{} level. The C-scheme additionally radiates partons, and they emphasize this is the theoretically preferred treatment.

Until now, \powhegbox{} did not include the FKS collinear remnant associated with the initial-state ${\gamma \rightarrow q \bar{q}}$ splittings. After \MSbar{} factorization of the corresponding singularity into the photon PDF, the finite remainder has the same form as the ${g \rightarrow q \bar{q} }$ splitting remnant, except for the replacement of $\alpha_S$ by $\alpha$ and the Altarelli-Parisi splitting function being
\begin{equation}
    P^{q \leftarrow \gamma}(z,\epsilon) = N_C e_q^2 \left[ 1- \frac{2z(1-z)}{1-\epsilon}\right]~.
\end{equation}
This differs from the splitting kernel $ P^{q \leftarrow g}$ only by replacing the color factor by the number of colors multiplied by the quark charge fraction squared, \ie~${T_F \rightarrow N_C e_q^2}$. We implemented this collinear remnant in \powhegbox{} arriving an implementation equivalent to the one in  \texttt{POWHEG\,BOX\,RES}~\cite{Jezo:2015aia}.

\subsection{\texttt{LHE} interface}
\label{sec: LHE interface}

The events generated by \powhegbox{} are stored in Les Houches Event (\texttt{LHE}) files~\cite{Alwall:2006yp} which are then passed to the shower. The \texttt{LHE} files contain the beam energies and the momenta of the incoming partons of the hard scattering, from which the parton shower can reconstruct the partons' momentum fractions. In resolved photoproduction, however, these only determine the parton-in-lepton momentum fraction, \ie~$x_A$ of Eq.~\eqref{eq:x-resolved}. Its split into the photon-in-lepton fraction $x_{\gamma/e}^{(A)}$ and the parton-in-photon momentum fraction $x_{i/\gamma}^{(A)}$ cannot be recovered. The decomposition could in principle be resampled by the shower according to the convolution of the photon flux and the photon PDFs at fixed $x_A$. This is consistent, however, only if the shower uses the same flux, PDF and factorization scale as \powhegbox. Moreover, such a resampling entails additional numerical effort, since it repeats a sampling that has already been performed in \powhegbox{}. Thus it is more efficient to pass the sampled value directly to the shower. To this end, for each event with a resolved photon, \powhegbox{} writes the sampled photon-in-lepton momentum fraction of the corresponding beam into the comment block of the event, in a line \texttt{\#xPhotonInBeam1} or \texttt{\#xPhotonInBeam2}, respectively. We extended the \texttt{LHE} reader of \pythia{}~\cite{Bierlich:2022pfr} to parse these extra comments and to use the values as the corresponding momentum fraction.

We have contacted the \pythia{} authors regarding an official implementation of this fix. They have recently provided us with a version based on the local modifications described above, which is not yet publicly available. We therefore expect this feature to be included in a future \pythia{} release, possibly together with an official version of the fix described below, which has not yet been addressed.

Since the resolved NLO process may also generate events with a photon in the initial state, as described in Sec.~\ref{sec: Modifications to powhegbox}, we add an event-by-event determination of the photon mode to \pythia{} for \texttt{LHE} input. For each event, the code checks whether or not the incoming parton on the lepton side is a direct photon or if it is resolved.

Although a complete description of the final state requires the \powhegbox{} events to be showered, it proves useful to analyse the \texttt{LHE} event output directly to examine the impact of the parton shower. Photoproduction analyses in \texttt{rivet}~\cite{Bierlich:2024vqo} usually reconstruct the photon kinematics from the momenta of the incoming and the scattered lepton. While this poses no problem for the showered events, since \pythia{} reconstructs the scattered lepton, the latter is not contained in the standard \texttt{LHE} file. Rather than modifying \texttt{rivet} to also read in the \texttt{\#xPhotonInBeam} comment, it showed to be more pragmatic to write out an additional auxiliary \texttt{LHE} file that contains the momentum of the scattered lepton $k^\prime=k-q$, where $k$ and $k^\prime$ denote the initial and final state lepton momenta and $q$ is the photon momentum. This auxiliary file is output in addition to the standard \texttt{LHE} file if the corresponding switch is enabled. It is used exclusively for the \texttt{LHE}-level analysis in this work and does not enter the showered predictions. Note, as the WWA photon flux is based on the collinear approximation and already integrated over the photon virtuality $Q^2$, the scattered lepton carries no transverse momentum in this setup. In the showered results, \pythia{} restores this information by resampling the $Q^2$ within the given bounds. However, the impact is negligible, since the scattering angle is small by construction due to the restricted photon virtuality.

\section{Numerical results}
\label{sec: results}

In the numerical results presented below, the \texttt{SaS1M} photon PDF~\cite{Schuler:1995fk,Schuler:1996fc} and the \texttt{CJ15nlo} proton PDF~\cite{Accardi:2016qay} are used unless indicated otherwise\footnote{Although \texttt{VALO10\_NLO\_MSbar} is a more recent \MSbar{} photon PDF set, we refrain from using it here, since at the time of writing the conversion from the \DISg\ to the \MSbar{} scheme in \texttt{VALO} was affected by an inconsistency.}, regardless of the perturbative accuracy of the prediction. Furthermore, we set the factorization scale of both the photon and the hadron equal to the renormalization scale with ${\mu_F^{(\gamma)}=\mu_F^{(H)}=\mu_R=H_T/2}$, where $H_T$ denotes the scalar sum of the transverse momenta of all final-state particles. 
Additionally, we set $1/\alpha=132.119$, as calculated from the Fermi constant and the $W$- and $Z$-boson mass in the $G_\mu$-scheme.
We shower the events using \texttt{PYTHIA\,8318} including the fixes described in Sec.~\ref{sec: LHE interface}. All event selection cuts and jet observables are implemented using \texttt{rivet\,4.1.3}~\cite{Bierlich:2024vqo}.

\subsection{Fixed-order validation}

The splitting of the incoming photon into a collinear quark-antiquark pair gives rise to an initial-state singularity, which consequently needs to be factorized in the photon PDF. This procedure introduces the unphysical factorization scale $\mu_F^{(\gamma)}$, which separates the direct from the resolved photon contributions. Up to higher order contributions the $\mu_F^{(\gamma)}$ dependence should cancel between the PDF and the collinear remnant of this splitting, as demonstrated in Refs.~\cite{Bodeker:1994fz,Klasen:1997br}. 

\begin{figure}
    \centering
    \includegraphics[width=0.55\linewidth]{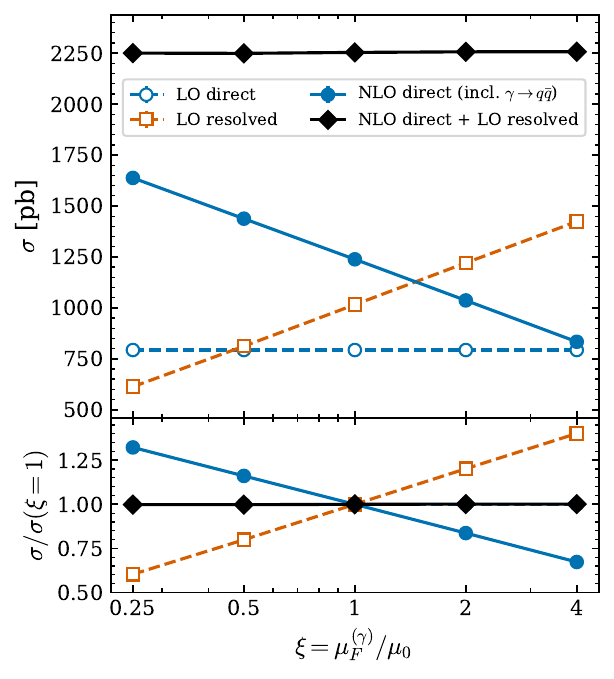}
    \caption{Dependence of the dijet photoproduction cross section on the photon factorization scale $\mu_F^{(\gamma)}=\xi \mu_0$ with $\mu_0=H_T/2$. The direct (blue) and resolved (orange) contributions are shown at LO (dashed), and the direct contribution is shown at NLO (solid). Note that here we explicitly add the ${\gamma \rightarrow q \bar{q}}$ splitting to it. The upper panel shows the total cross section and the lower panel shows the cross section ratio with respect to ${\xi=1}$.}
    \label{fig: photon factorization scale dependence}
\end{figure}

Figure~\ref{fig: photon factorization scale dependence} shows the dependence of the total cross section for $e^+p$ collisions on this scale. In this figure, we set $\mu_F^{(\gamma)}=\xi \mu_0$ with $\mu_0= H_T/2$, where $\xi$ is a variation factor, while the proton factorization scale is kept at $\mu_F^{(P)}=\mu_0$. Furthermore, the beam energies mimic a \hera{1} like setup, $E_e=\SI{27.5}{\giga \electronvolt}$ and $E_p=\SI{820}{\giga \electronvolt}$, and the jet finder requires $|\eta|\leq 4$ and imposes a leading (subleading) jet cut of $p_T\geq \SI{14}{\giga \electronvolt}$ ($p_T\geq \SI{11}{\giga \electronvolt}$). Since the parton distribution of the photon in the photon is formally given by $\delta(1-x_{i/\gamma})$, the direct LO contribution does not depend on the factorization scale, as confirmed by the dashed blue line. On the other hand, the resolved LO contribution scales logarithmically with $\xi$, demonstrating the photon PDF dependence. Following the explanation in Sec.~\ref{sec: Modifications to powhegbox}, the $\gamma \rightarrow q \bar{q}$ splitting is normally associated with the NLO resolved contribution in our implementation. However, in order to showcase its impact, we exclude it here from the NLO resolved contribution and add it, together with all other QED singular regions built on top of resolved Born contributions, to the NLO direct contribution. Following this, the solid blue NLO direct contribution exhibits a negatively logarithmic dependence, nearly completely canceling the photon PDF scale dependence as can be seen by the black line. The behavior under variations of $\mu_F^{(\gamma)}$ therefore provides a valuable consistency check of our implementation.

Monte Carlo event generators are predominantly developed and validated for symmetric collider setups like LHC or Tevatron. Although \powhegbox{} is designed keeping in mind asymmetric collisions such as $e^+p$ scattering, this functionality has been tested far less extensively. A simple test is provided by exchanging the two beams. Any observable that is invariant under this exchange must yield identical predictions for $e^+p$ and $pe^+$ collisions, whereas a discrepancy would indicate an inconsistent treatment of the beams.
\begin{figure}
    \centering
    \begin{subfigure}{0.49 \textwidth}
        \centering
        \includegraphics[width=\linewidth]{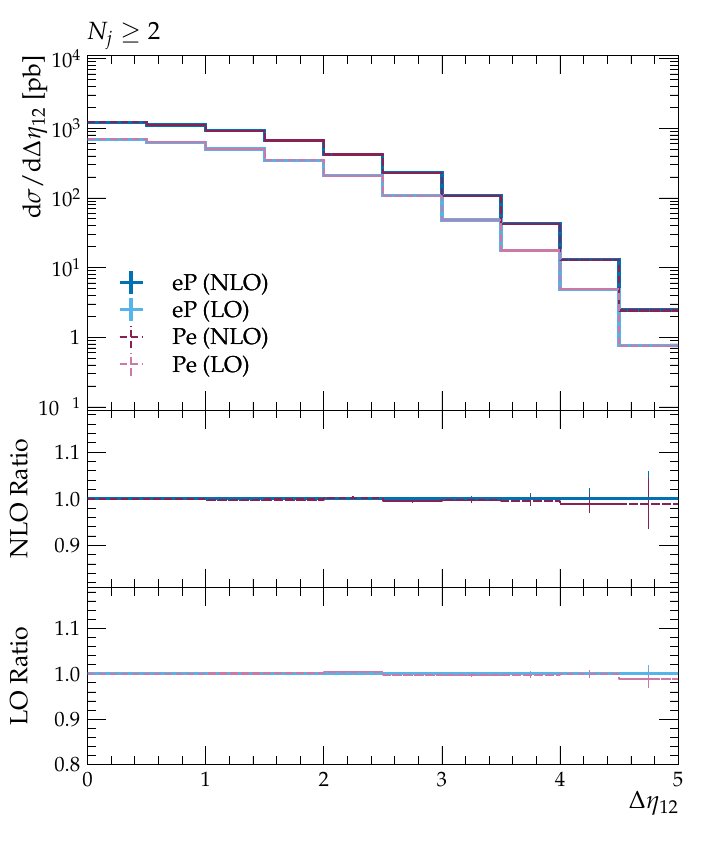}
    \end{subfigure}
    \begin{subfigure}{0.49 \textwidth}
        \centering
        \includegraphics[width=\linewidth]{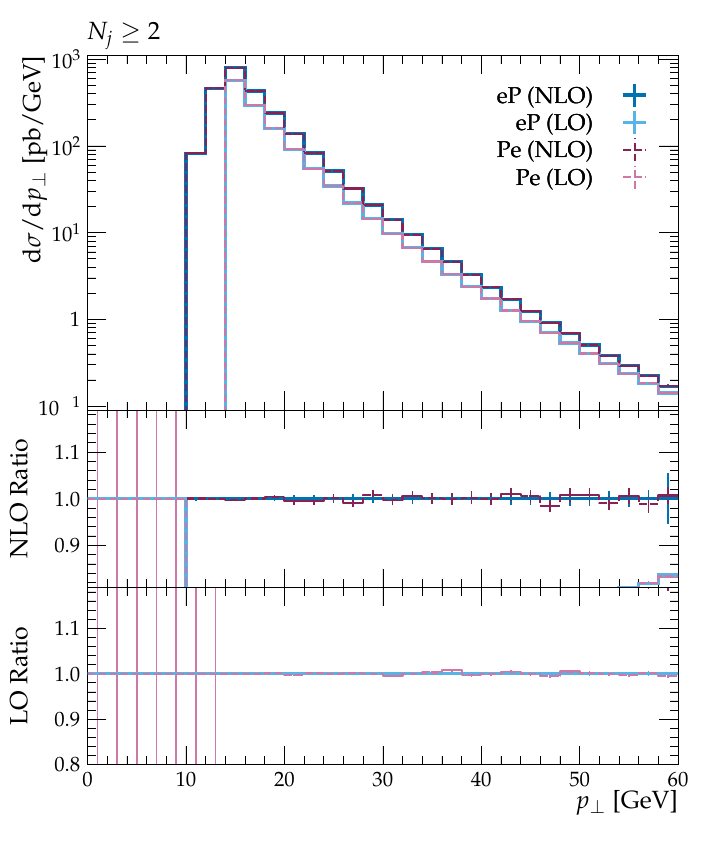}
    \end{subfigure}
    \caption{Fixed-order differential cross sections as functions of $\Delta \eta_{12}$ (left) and $p_T$ (right). The results are shown for $e^+p$ and $pe^+$ collisions. For $e^+p$, NLO is shown in a darker and LO in lighter shade of blue; for $pe^+$, NLO is shown in a darker and LO in a lighter shade of violet. The middle panels show the ratio of the NLO result, while the bottom panels show the LO ratios.}
    \label{fig: fixed-order-symmetry}
\end{figure}
To test this, we again consider $E_e=\SI{27.5}{\giga \electronvolt}$ and $E_p=\SI{820}{\giga \electronvolt}$ with jet cuts of $|\eta|\leq 4$ and leading (subleading) jet transverse momenta of $p_T\geq \SI{14}{\giga \electronvolt}$ ($p_T\geq \SI{11}{\giga \electronvolt}$). In contrast to the previous discussion of the factorization scale dependence, we now fix the renormalization and factorization scales to the $Z$-boson mass, $\mu=m_Z=\SI{91.1876}{\giga \electronvolt}$, and use running $\alpha_S$ evolved by LHAPDF using the prescription shipped with the proton PDF set to ensure that both configurations are treated with exactly the same theoretical input. Since the cuts are symmetric in $\eta$, the absolute pseudorapidity difference between the two leading jets, $\Delta\eta_{12}=|\eta_1-\eta_2|$, and the inclusive jet $p_T$ are both invariant under beam exchange. Figure~\ref{fig: fixed-order-symmetry} shows the corresponding LO and NLO fixed-order predictions. The different impact of the subleading $p_T$ cut at LO and NLO arises because the two LO final-state jets are always produced back-to-back. If one jet satisfies the leading $p_T$ cut, the other automatically satisfies it as well. At NLO, additional emissions and the corresponding recoil make the subleading jet cut physically relevant.
The top row of Fig.~\ref{fig: fixed-order-symmetry} already indicates good agreement between the two beam-exchanged setups. The middle panels show in more detail at NLO that both setups agree within numerical uncertainties with the central predictions differing by less than $\SI{1}{\percent}$. The bottom panels display the corresponding LO ratios, which also show sub-percent agreement, confirming that asymmetric beam configurations are treated consistently in our implementation.

\begin{figure}
    \centering
    \begin{subfigure}{0.49 \textwidth}
        \centering
        \includegraphics[width=\linewidth]{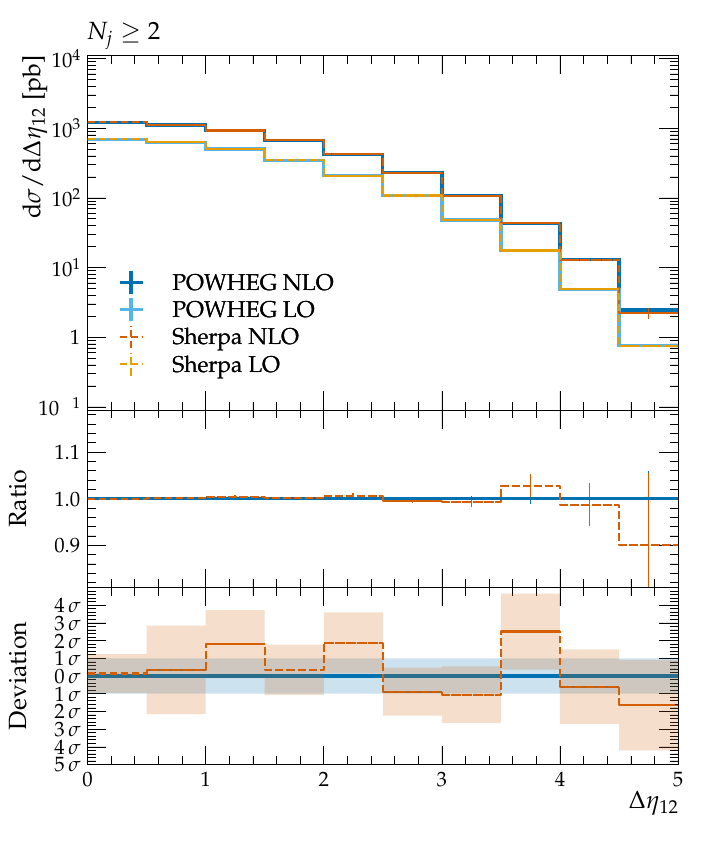}
    \end{subfigure}
    \begin{subfigure}{0.49 \textwidth}
        \centering
        \includegraphics[width=\linewidth]{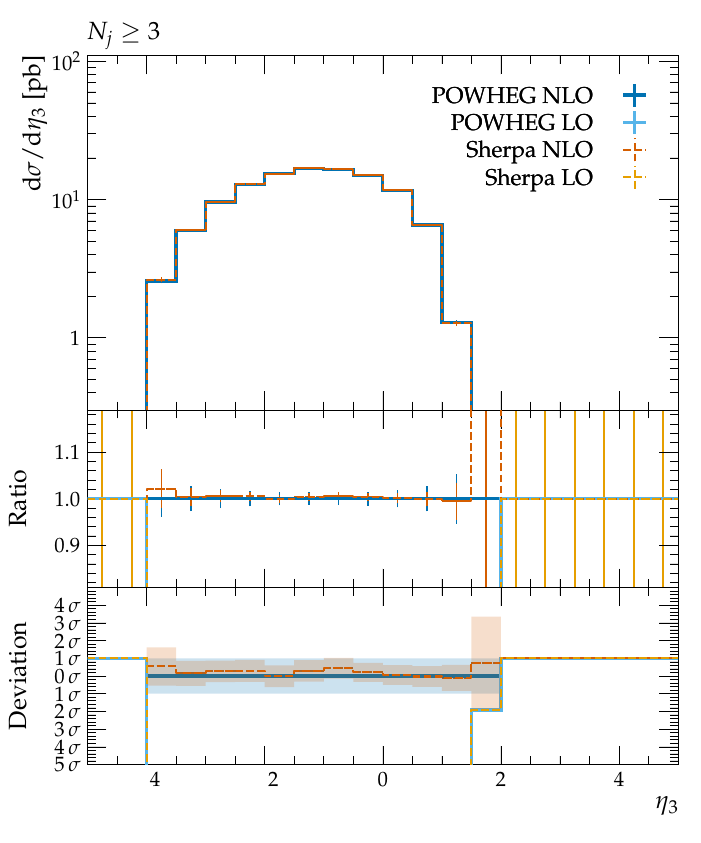}
    \end{subfigure}
    \begin{subfigure}{0.49 \textwidth}
        \centering
        \includegraphics[width=\linewidth]{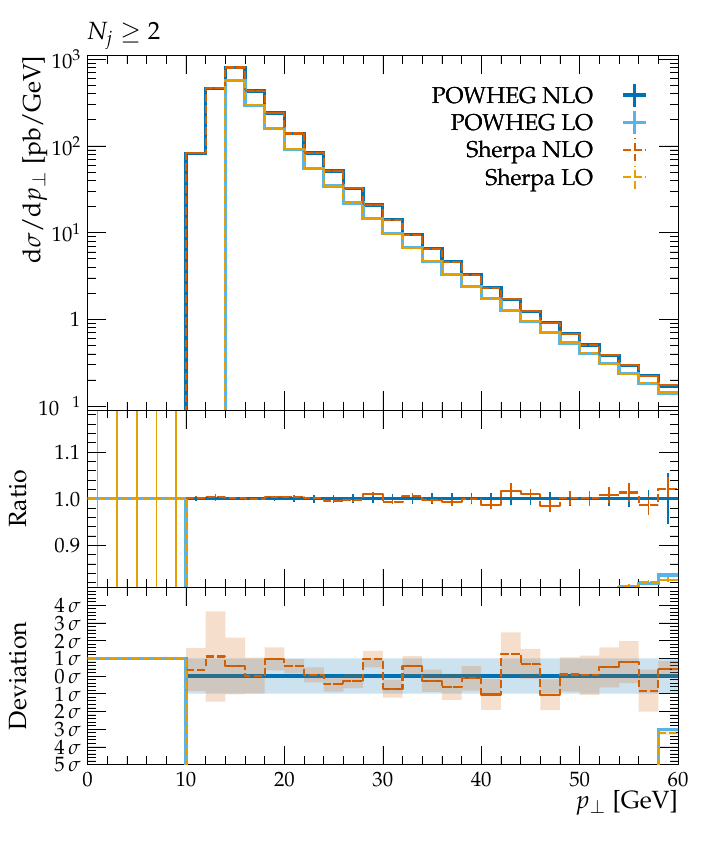}
    \end{subfigure}
    \begin{subfigure}{0.49 \textwidth}
        \centering
        \includegraphics[width=\linewidth]{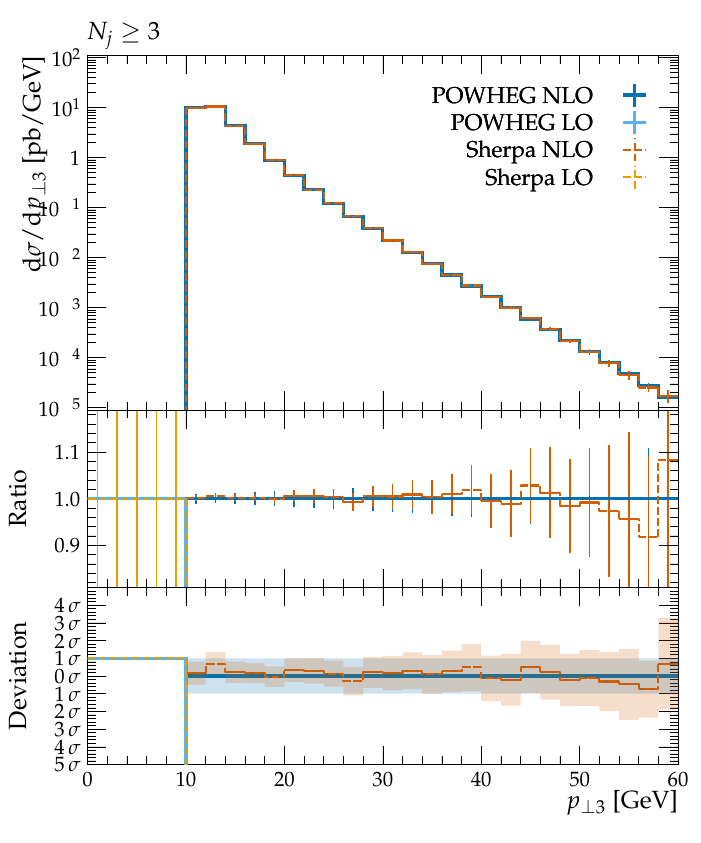}
    \end{subfigure}
    \caption{Fixed-order differential cross sections as functions of $\Delta \eta_{12}$ (top, left), $\eta_3$ (top, right), $p_T$ (bottom, left), and $p_{T,3}$ (bottom, right). The lines show the NLO and LO results from \powhegbox{} and \sherpa. For \powhegbox{}, NLO is shown in a darker and LO in lighter shade of blue; for \sherpa{}, NLO is shown in a darker and LO in a lighter shade of orange. The middle panels show the ratios of the NLO results, while the bottom panels quantify their deviations. Numerical errors are shown as error bars in the top and middle panels and as error bands in the lower panels.}
    \label{fig: fixed-order-pwg-sherpa}
\end{figure}

Finally we also compare our predictions to those obtained with a patched version of SHERPA \cite{Hoeche:2023gme,Sherpa:2024mfk,PrivateCommunicationPM}.
It is important to stress that at LO the direct and resolved contributions are expected and are also found to agree separately between the two generators. At NLO, however, only their sum is meaningful for this comparison, because the prescription for the separation into a direct and a resolved contribution does not coincide between \powhegbox{} (see Sec.~\ref{sec: Modifications to powhegbox}) and \sherpa. For this reason, in Fig.~\ref{fig: fixed-order-pwg-sherpa}, we compare only the sum of both contributions for fixed-order distributions at both NLO and LO as obtained from \powhegbox{} and \sherpa.

On the left panels of this figure, we show the differential cross sections as functions of $\Delta\eta_{12}$ and $p_T$. At NLO, the two predictions have overlapping numerical errors and can therefore be considered to be in agreement. Although we do not display an explicit ratio plot, the same level of agreement is also observed at LO. Since these are two-jet inclusive observables, they provide a global test including a test of the treatment of IR singularities, which differs between the two tools. In order to probe this part of the phase space in a more exclusive way, we additionally consider observables that are sensitive to the third jet, namely its pseudorapidity $\eta_3$ and transverse momentum $p_{T,3}$ on the right-hand side of the figure. For these observables, the two generators agree as well within numerical uncertainties. Given the different technical implementations in \powhegbox{} and \sherpa{}, their agreement constitutes an important further step in the validation of our implementation.

\subsection{Comparison to HERA results}
\label{sec: Comparison to HERA Results}

Having concluded the discussion of our validation studies, we now turn to the phenomenological predictions.
Between 1992 and 2007, collisions of electrons and positrons with protons were produced at HERA, allowing the investigation of photoproduction processes. During its first phase of operation, \hera{1}, the beam energies were set to $E_e=\SI{27.5}{\giga \electronvolt}$ and $E_p=\SI{820}{\giga \electronvolt}$. After its upgrade, \hera{2} ran from 2003 at higher luminosities and increased proton energy $E_p=\SI{920}{\giga \electronvolt}$~\cite{Klein:2008di}. We focus on two jet photoproduction analyses by the ZEUS Collaboration~\cite{ZEUS:2001zoq,ZEUS:2007njl}, based on data taken during \hera{1} and \hera{2}, respectively. To mirror the experimental analysis cuts of $E_T^{\text{jet1}}>\SI{14}{\giga \electronvolt}$ ($E_T^{\text{jet1}}>\SI{20}{\giga \electronvolt}$) for the leading jet and $E_T^{\text{jet2}}>\SI{11}{\giga \electronvolt}$ ($E_T^{\text{jet2}}>\SI{15}{\giga \electronvolt}$) for the subleading jet in the Run-1 (Run-2) analysis, we set the underlying Born transverse momentum cut to $k_T^{\text{min}}=\SI{8}{\giga \electronvolt}$ ($k_T^{\text{min}}=\SI{10}{\giga \electronvolt}$) and the maximal photon virtuality to $Q^2=\SI{1}{\giga \electronvolt^2}$.

The \hera{1} analysis~\cite{ZEUS:2001zoq} uses the absolute value of the observable
\begin{align}
    \cos \theta^* = \tanh \left( \frac{\eta^{\text{jet1}}-\eta^{\text{jet2}}}{2} \right) ~,
\end{align}
where $\eta^{\text{jet1}}$ and $\eta^{\text{jet2}}$ are the pseudorapidities of the two jets with the highest transverse energy. As $\theta^*$ reflects the scattering angle in the partonic center-of-mass frame, it is directly sensitive to the hard-scattering matrix element, while also being invariant under longitudinal boosts. Additionally, the analysis introduced $x_{\gamma}^{\text{obs}}$, an observable estimating the fraction of the photon momentum carried by the parton entering the hard scattering, 
\begin{align}
    x_{\gamma}^{\text{obs}}= \frac{E_T^{\text{jet1}} e^{-\eta^{\text{jet1}}} + E_T^{\text{jet2}} e^{-\eta^{\text{jet2}}}}{2 x_{\gamma/e} E_e}~, \label{eq. x_gamma_obs}
\end{align}
where $E_e$ is the lepton beam energy.
At LO $x_{\gamma}^{\text{obs}}$ coincides with the true momentum fraction previously denoted as $x_{i/\gamma}$.

\subsubsection{Impact of the parton shower}
\label{sec: Impact of the parton shower}

\begin{figure}
    \centering
    \begin{subfigure}{0.49 \textwidth}
        \centering
        \includegraphics[width=\linewidth]{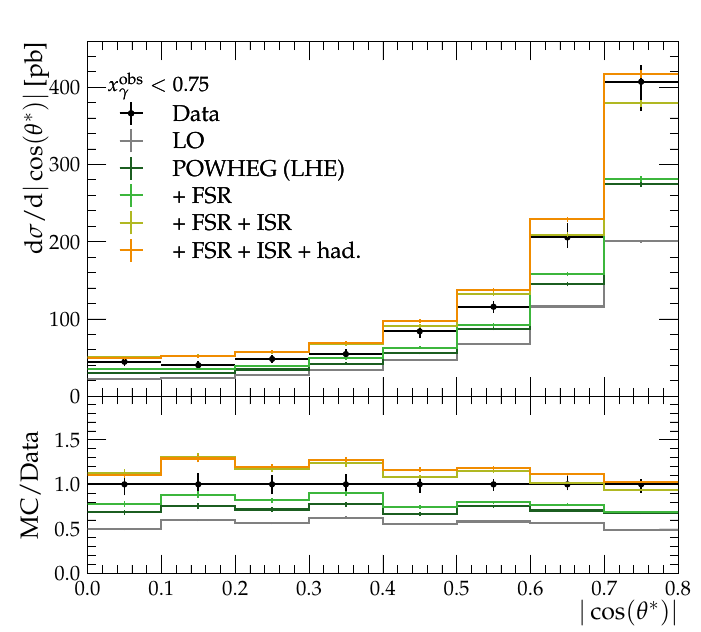}
    \end{subfigure}
    \begin{subfigure}{0.49 \textwidth}
        \centering
        \includegraphics[width=\linewidth]{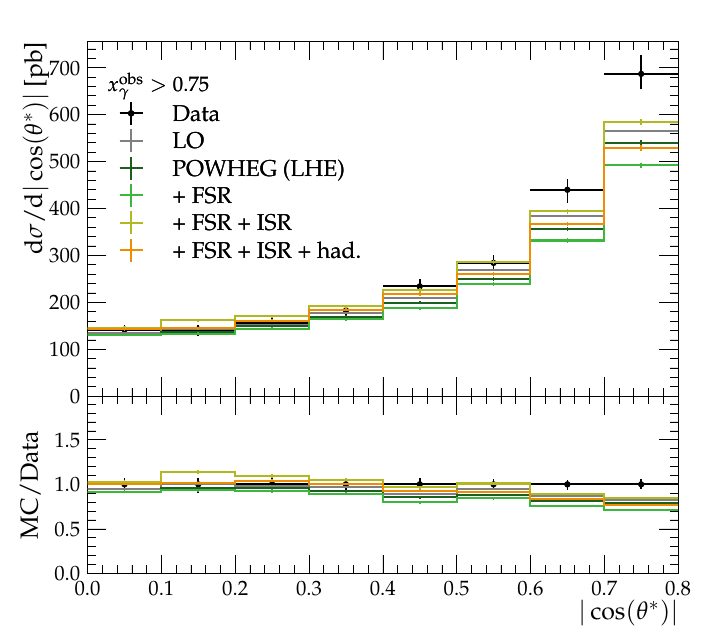}
    \end{subfigure}
    \caption{Differential dijet photoproduction cross section as a function of $|\cos \theta^*|$ for $x_{\gamma}^{\text{obs}}\leq 0.75$ (left) and $x_{\gamma}^{\text{obs}}> 0.75$ (right) as measured in \hera{1} by ZEUS~\cite{ZEUS:2001zoq}. It is compared to LO fixed-order (grey) and NLO predictions generated with \powheg{} and matched to the \pythia{} parton shower, shown at successive stages: at \texttt{LHE}-level (dark green), adding final-state radiation only (light green), adding initial-state radiation (olive) and adding hadronization (orange). Numerical uncertainties are indicated by error bars.}
    \label{fig:HERA-Run-1-shower}
\end{figure}

In dijet photoproduction, the parton shower can induce sizable corrections in parts of the fiducial phase space. In this section we will illustrate these effects using the ZEUS~\cite{ZEUS:2001zoq} dijet photoproduction analysis. At the same time, this study fixes the shower settings used throughout the remainder of this work. We do not consider multi-parton interactions (MPI), as they are not related to the NLO matching and merit a separate analysis. We also note that no explicit $ep$ shower tune is available for photoproduction, with existing efforts focusing on constraining the MPI parameters~\cite{Butterworth:2024hvb}. If not explicitly stated otherwise, we adopt the default settings. For comparison, we also show \texttt{LHE}-level results alongside the showered predictions, obtained from the auxillary \texttt{LHE} file described in Sec.~\ref{sec: LHE interface}.

In Fig.~\ref{fig:HERA-Run-1-shower}, we show the differential cross section with respect to the scattering angle $\theta^*$, divided into a resolved-dominated part, defined by $x_{\gamma}^{\text{obs}}<0.75$ (left), and a direct-dominated region, given by $x_{\gamma}^{\text{obs}}>0.75$ (right). Comparing the LO fixed-order (grey) and the NLO \texttt{LHE}-level results (dark green), we find a $K$-factor of approximately $1.4$ in the resolved-dominated region. This factor stays largely constant, \ie~the NLO corrections do not alter the shape of the distribution. At the full NLO+PS level (orange), we again observe no change in shape, with a ratio of about $1.6$ relative to the NLO \texttt{LHE}-level result. In the direct-dominated region, the corrections are less pronounced, with ratios close to unity. Apart from the large $K$-factors, we thus find that in the current fiducial region the higher order corrections are in general well behaved.

To better understand the origin of the enhancement at the NLO+PS level, we now inspect the individual contributions from initial-state radiation (ISR), final-state radiation (FSR) and hadronisation added subsequently to the NLO \texttt{LHE}-level results. Since the shower corrections are predominantly flat they can be explained as migration over the fiducial cut boundary. Even though the impact of the shower is unitary at the inclusive level, which we have verified explicitly, the fiducial cross section is suppressed by about one order of magnitude, thus a relatively small shape correction to the spectra used for the cut can lead to a relatively large normalization correction. Indeed, the fiducial cross section integrated over $x_{\gamma}^{\text{obs}}$ receives a NLO+PS/LHE correction of about \SI{32}{\percent}. 

The impact of the different contributions of the parton shower is more prominent in the resolved-dominated region in Fig.~\ref{fig:HERA-Run-1-shower}, which can be better understood as follows. A direct event, characterized by $x_{\gamma}^{\text{obs}}\approx1$, can radiate a hard jet during the parton shower. The recoil from this reduces the momenta of the other jets, thereby decreasing the value of $x_{\gamma}^{\text{obs}}$ as defined in Eq.~\eqref{eq. x_gamma_obs}. If this decrease is sufficiently large, or the event was already close to the threshold of $x_{\gamma}^{\text{obs}}=0.75$, it might fall below. In addition to this migration from the direct to the resolved region, the parton shower only adds initial-state radiation to the protons and resolved photons, yielding a more pronounced effect in the resolved region as opposed to the direct-dominated region. A reassuring observation is that the parton shower corrections are fairly flat across both regions, indicating that the shower does not introduce any significant shape distortion. The largest impact comes from initial-state radiation in the resolved-dominated region, which yields a correction factor of about $1.5$. Final-state radiation and hadronization also contribute by a factor of about $1.2$, but are subleading. 

Given this, it is worthwhile to point out differences in our treatment of the initial- and final-state showers. As discussed in Sec.~\ref{sec: Modifications to powhegbox}, we do not want the shower to generate emissions for scales that were not evolved in \powhegbox. Since in \powhegbox{} the QED scale is only evolved for the initial-state $\gamma \to q \bar{q}$ splitting, through which the backward evolution of a quark terminates on a direct beam photon, it would be ideal to add only these QED splittings in the parton shower. However, in \pythia{} this splitting is controlled by the same switch as photon emission off incoming quarks, \texttt{SpaceShower:QEDshowerByQ}, and therefore cannot be enabled independently. Consequently, we switch off \texttt{QEDshowerByQ} and \texttt{QEDshowerByL} for the final-state and the initial-state shower added to both contributions, with the exception of keeping \texttt{QEDshowerByQ=on} for the initial-state shower attached to the resolved contribution. In the backward evolution of the initial-state shower, \pythia{} reweights the splitting probability by a ratio of the corresponding PDF. For the proton, the PDF set can be chosen with the option \texttt{PDF:pSet}. For the photon, however, there is no such choice and \pythia{} always uses the CJKL set~\cite{Cornet:2002iy}. This introduces a mild inconsistency between the hard generation and the shower. Using a different photon PDF set, such as \texttt{VALO}, in the shower is, however, not expected to yield large changes~\cite{PrivateCommunicationIH}.

\begin{figure}
    \centering
    \includegraphics[width=0.55\linewidth]{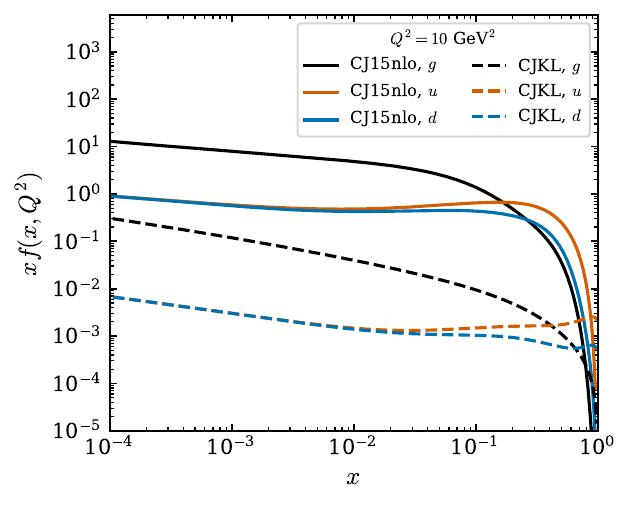}
    \caption{Parton distributions of the proton and the photon at $Q^2=\SI{10}{\giga \electronvolt}^2$ for the gluon (black), up quark (orange) and down quark (blue). Solid lines show the central values of the CJ15nlo proton PDFs and dashed lines show the CJKL photon PDFs.}
    \label{fig: CJ15_vs_CJKL}
\end{figure}

A more relevant aspect is the behavior of the PDF ratio in backward evolution itself. The probability that parton $b$ originates from a splitting off parton $a$, $d\mathcal{P}_b$, is proportional to the PDF ratio as~\cite{Bierlich:2022pfr}
\begin{align}
    d\mathcal{P}_b \propto \sum\limits_a\int dz \frac{\frac{x}{z}f_a(\frac{x}{z}, Q^2)}{xf_b(x,Q^2)} P_{a\to b}(z),
\end{align}
where $x$ is the momentum fraction of parton $b$ and $z$ is the momentum fraction carried by $b$ with respect to its parent $a$. Since $z<1$, the PDFs are evaluated at increasingly larger momentum fractions, $x/z>x$. For the proton, the PDF falls steeply as $x\to 1$, so the ratio is small and backward splittings become suppressed. For the photon, in contrast, the quark content does not fall as steeply at large $x$ as shown in Fig.~\ref{fig: CJ15_vs_CJKL}, leading to a larger ratio and hence a higher splitting probability. Since \pythia{} does not allow the initial-state shower to be enabled independently for the two beams, this hypothesis cannot be tested directly. Nevertheless, it provides a plausible explanation for the large increase induced by the initial-state shower.

The interplay between \powhegbox{} and \pythia{} is governed by a veto procedure with the parton shower starting at the kinematic limit (\texttt{pTmaxMatch = 2}), and every shower emission harder than the \powhegbox{} emission is vetoed (\texttt{veto = 1}). The definition of the veto scale, the computation of the hardness of each shower emission, and the veto itself are implemented in the \hooks{} header. A common choice adopted throughout this paper is to take the veto scale directly from the \texttt{SCALUP} value of the \texttt{LHE} event (\texttt{pThard = 0}), to compute the emission hardness with respect to the radiating parton (\texttt{pTemt = 0}), and to use the \powhegbox{} definition for hardness, \ie~the transverse momentum for ISR and the distance measure $d_{ij}$ for FSR, in \pythia{} (\texttt{pTdef = 1}). We note, however, that there is no consensus on the optimal settings, and variations of these flags can provide an estimate of the associated matching uncertainty. A study of such variations in dijet hadroproduction~\cite{Buckley:2016bhy} found that \texttt{pTemt = 0} is the only satisfactory choice, whereas no preferred setting was found for the remaining parameters\footnote{Note, however, that the \texttt{pThard = 0} option was not considered in this study.}.

\begin{figure}
    \centering
    \begin{subfigure}{0.49 \textwidth}
        \centering
        \includegraphics[width=\linewidth]{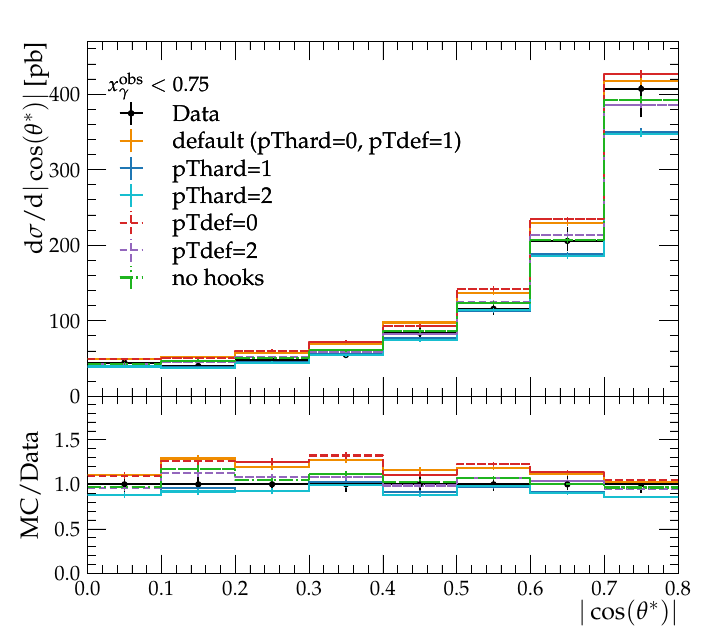}
    \end{subfigure}
    \begin{subfigure}{0.49 \textwidth}
        \centering
        \includegraphics[width=\linewidth]{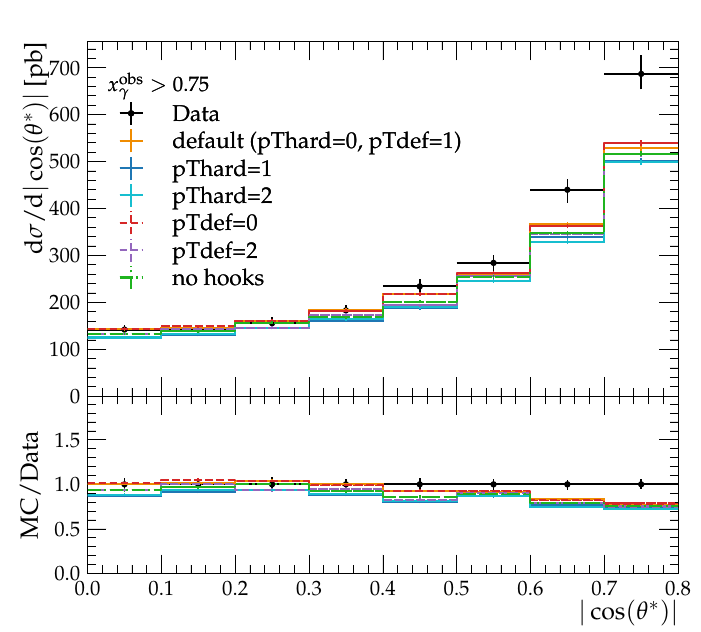}
    \end{subfigure}
    \caption{Differential dijet photoproduction cross section as a function of $|\cos \theta^*|$ for $x_{\gamma}^{\text{obs}}\leq 0.75$ (left) and $x_{\gamma}^{\text{obs}}> 0.75$ (right) as measured in \hera{1} by ZEUS~\cite{ZEUS:2001zoq}. Shown are NLO+PS predictions with varied \hooks{} settings. Our default \hooks{} choice (solid, orange) is compared against variations of \texttt{POWHEG:pThard} (solid, blue and cyan), variations of \texttt{POWHEG:pTdef} (dashed, red and violet), and no usage of \hooks{} (dashed, blue). Numerical uncertainties are indicated by error bars. The ratios with respect to our default \hooks{} choice are shown in the lower panels.}
    \label{fig:HERA-Run-1-hooks}
\end{figure}

In Fig.~\ref{fig:HERA-Run-1-hooks} we show the impact of varying \texttt{pThard} and \texttt{pTdef} individually with respect to our default choice, which does not coincide with the \pythia{} default settings. Instead of using \texttt{SCALUP} directly, the veto scale can be recomputed from the event kinematics, either as the minimal relative transverse momentum of the \powhegbox{} emission with respect to all other partons (\texttt{pThard=1}), or as the minimal relative transverse momentum of any final-state parton with respect to all other partons (\texttt{pThard=2}). Since the minimum is taken, the latter yields the lower veto scale and hence the stronger restriction of the shower phase space. Both options reduce the fiducial cross section by about \SI{15}{\percent}, reflecting that the scale between emitter and emitted parton as calculated in \powhegbox{} yields higher values than between other parton pairs in the event. The impact is especially visible in the shower-sensitive resolved-dominated region in Fig.~\ref{fig:HERA-Run-1-hooks}, with the reduction reaching about \SI{30}{\percent}. When switching the scale definitions, using the \powhegbox{} ISR transverse momentum also for FSR (\texttt{pTdef=0}) leaves the fiducial cross section nearly unchanged, and the deviations in the resolved-dominated region stay within \SI{5}{\percent}, mostly compatible with the statistical uncertainty. We can therefore deduct that ISR and FSR scale definitions in \powhegbox{} yield similar results. Using the \pythia{} definitions (\texttt{pTdef=2}) on the other hand reduces the fiducial cross section by about \SI{10}{\percent}, and by up to \SI{20}{\percent} in the resolved-dominated region. For completeness, Fig.~\ref{fig:HERA-Run-1-hooks} also shows the result without the \hooks{} veto procedure, with the shower limited to the scale of the \texttt{LHE} event (\texttt{pTmaxMatch = 1} and \texttt{veto = 0}). It showcases that the impact of the \hooks{} yields a correction of about \SI{20}{\percent} for the resolved-dominated results.

Summarizing, the variations in the shower matching procedure are most sensitive to the hardness of the \texttt{LHE} event, resulting in a \SI{15}{\percent} reduction over the fiducial cuts and a \SI{30}{\percent} reduction in the resolved-dominated region. Keeping this matching uncertainty in mind, we continue by comparing to data. An example of the \pythia{} settings card used throughout this work and consistently shown with solid orange lines in Figs.~\ref{fig:HERA-Run-1-shower} and~\ref{fig:HERA-Run-1-hooks} is given in App.~\ref{sec: appendix Pythia8 settings}.

\subsubsection{\hera{1}}

\begin{figure}
    \centering
    \begin{subfigure}{0.49 \textwidth}
        \centering
        \includegraphics[width=\linewidth]{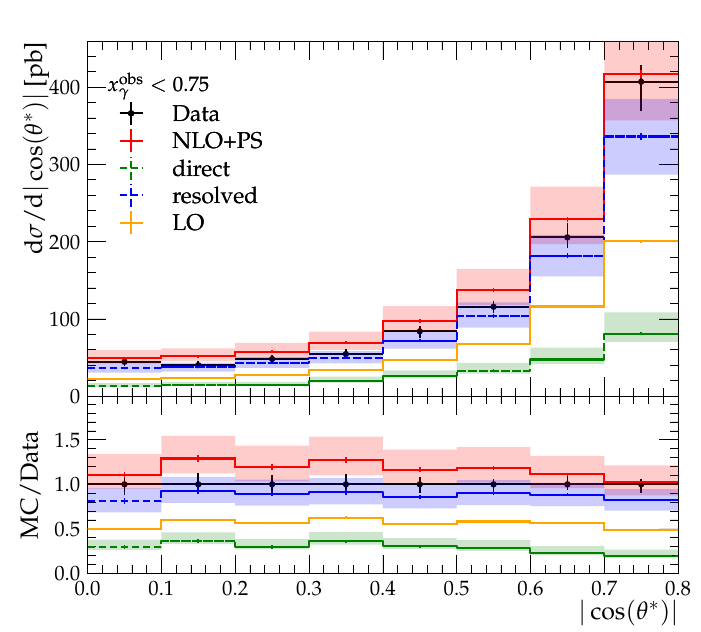}
    \end{subfigure}
    \begin{subfigure}{0.49 \textwidth}
        \centering
        \includegraphics[width=\linewidth]{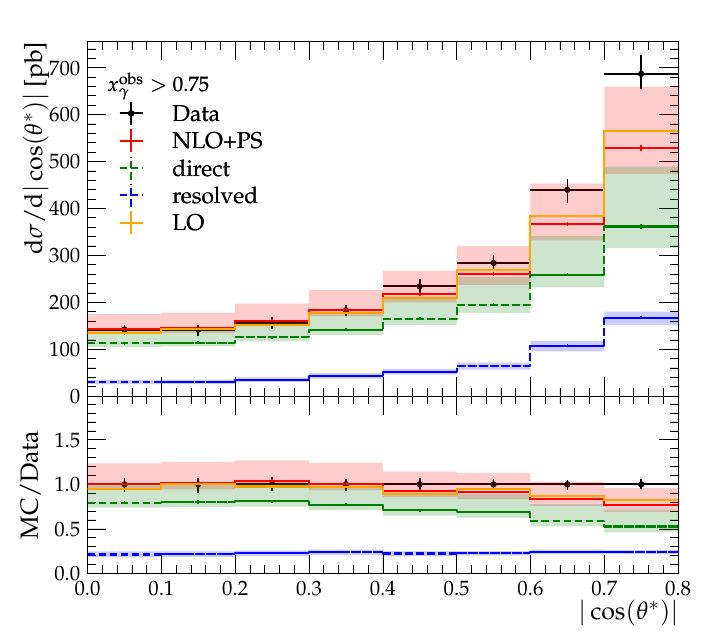}
    \end{subfigure}
    \begin{subfigure}{0.49 \textwidth}
        \centering
        \includegraphics[width=\linewidth]{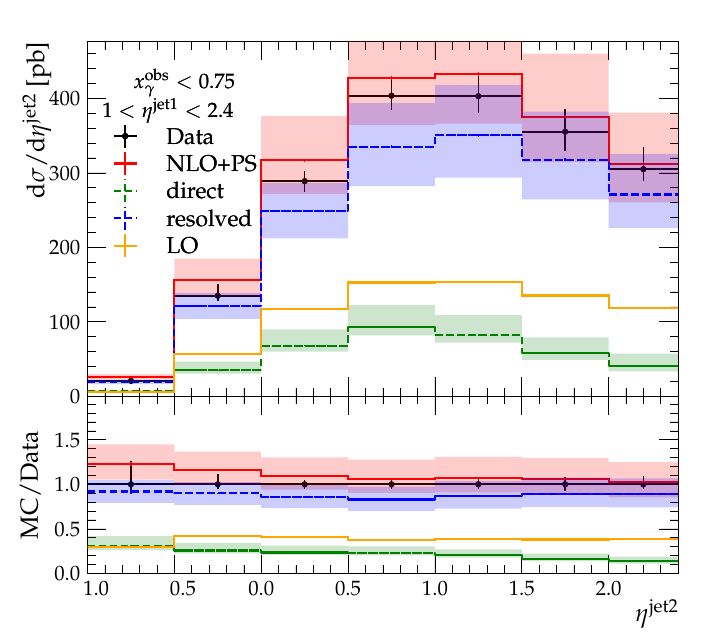}
    \end{subfigure}
    \begin{subfigure}{0.49 \textwidth}
        \centering
        \includegraphics[width=\linewidth]{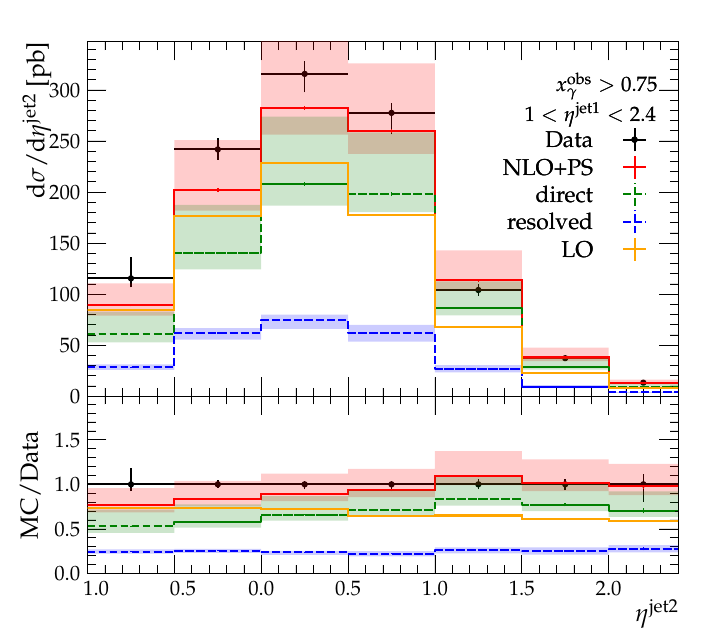}
    \end{subfigure}
    \caption{Differential dijet photoproduction cross section as a function of $|\cos \theta^*|$ (top) and $\eta^{\text{jet2}}$ (bottom) for $x_{\gamma}^{\text{obs}}\leq 0.75$ (left) and $x_{\gamma}^{\text{obs}}> 0.75$ (right) as measured in \hera{1} by ZEUS~\cite{ZEUS:2001zoq}. It is compared to NLO+PS predictions from the \powhegbox{} matched to a \pythia{} parton shower. The direct and resolved contributions are shown in green and blue, respectively, and their sum in red. Additionally the LO fixed-order sum is shown in yellow. Shaded bands indicate the envelope of a seven-point variation of the renormalization and factorization scales, while error bars indicate the numerical uncertainty.}
    \label{fig:HERA-Run-1-SaS1M}
\end{figure}

In Fig.~\ref{fig:HERA-Run-1-SaS1M} we show the \powhegbox{} NLO predictions matched to the \pythia{} parton shower, using the setup introduced in Sec.~\ref{sec: Comparison to HERA Results} with the \texttt{SaS1M} photon PDF and the \texttt{CJ15nlo} proton PDF. In addition to the full NLO+PS result, the direct and resolved contributions are shown separately. We note again that only the sum provides a physical prediction. All NLO+PS curves include the uncertainty from the standard seven-point variation of the factorization and renormalization scales. For comparison, we show the central fixed-order LO prediction for the sum of both contributions. The upper panels show the cross section differential in the cosine of the center-of-mass scattering angle in the resolved-dominated ($x_{\gamma}^{\text{obs}}<0.75$) and the direct-dominated ($x_{\gamma}^{\text{obs}}>0.75$) region. The fact that the individual contributions are dominant in only one of those regions supports the chosen thresholds, as one would expect. At LO, the predictions already reproduce the shape characteristics of the data while undershooting by a factor of two in the resolved region and partially agreeing in the direct region. At NLO+PS, the prediction in the resolved-dominated region moves considerably closer to the data, the changes in the direct region being rather minimal. In both cases the final prediction agrees with the data within the scale variation band in most bins. The scale uncertainty of about \SI{20}{\percent} is larger than the experimental uncertainty. Further sources of theoretical uncertainty, in particular the proton PDF variation, but also the photon PDF variation, and the matching settings discussed in Sec.~\ref{sec: Modifications to powhegbox} are not included in this band. However, even without including them, we find satisfactory agreement between data and predictions for these observables. The lower panels of Fig.~\ref{fig:HERA-Run-1-SaS1M} show the cross section differential in the pseudorapidity of the subleading jet, $\eta_2$, in the same two regions, while cutting also on the pseudorapidity of the first jet. Both the LO and the NLO+PS predictions reproduce the shape of the data, and the NLO+PS result agrees with the data within scale-variation uncertainties.

\subsubsection{\hera{2}}
\label{sec: hera2}

\begin{figure}
    \centering
    \begin{subfigure}{0.32 \textwidth}
        \centering
        \includegraphics[width=\linewidth]{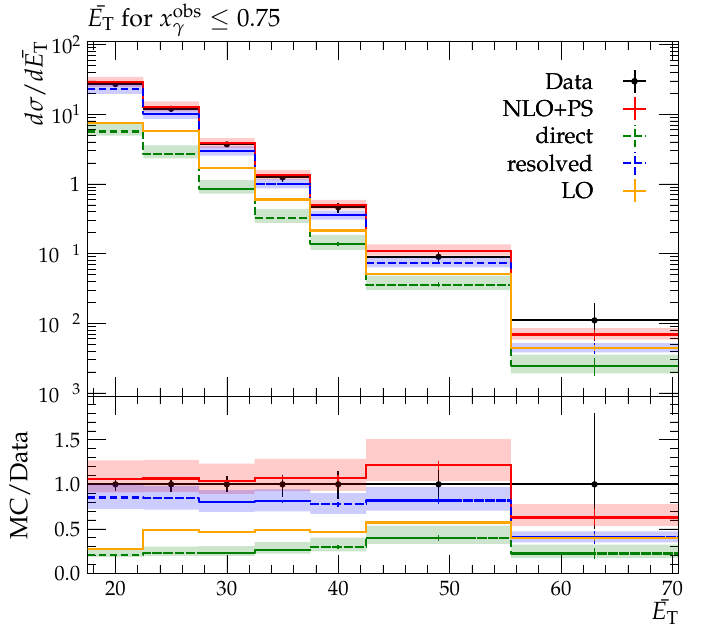}
    \end{subfigure}
    \begin{subfigure}{0.32 \textwidth}
        \centering
        \includegraphics[width=\linewidth]{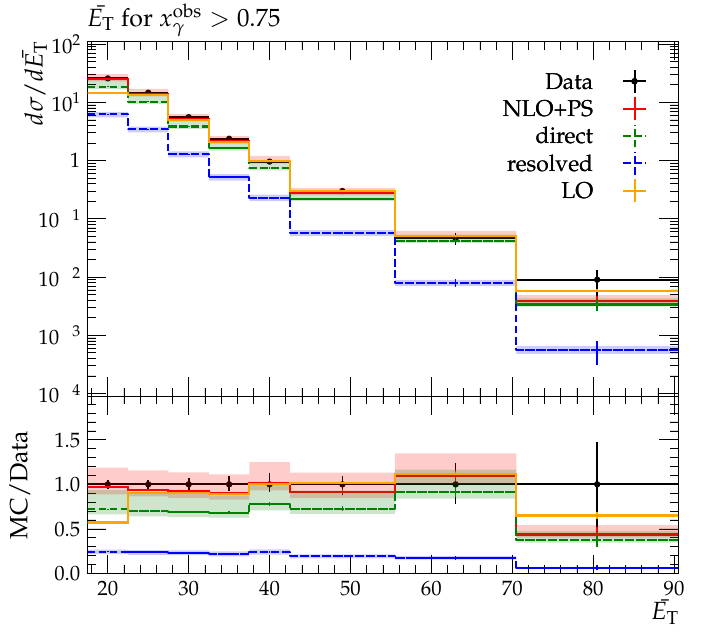}
    \end{subfigure}
    \begin{subfigure}{0.32 \textwidth}
        \centering
        \includegraphics[width=\linewidth]{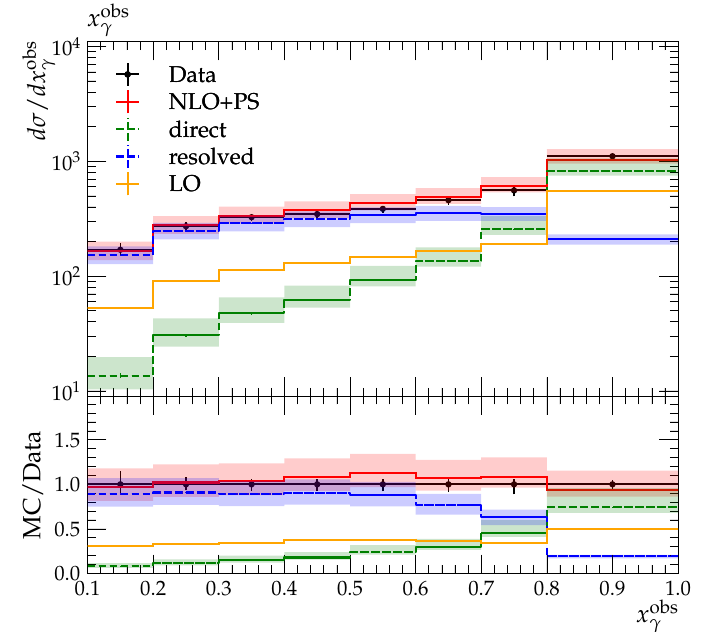}
    \end{subfigure}
    \caption{Dijet photoproduction cross section differential in $\bar{E}_T$ for $x_{\gamma}^{\text{obs}}\leq 0.75$ (left) and $x_{\gamma}^{\text{obs}}> 0.75$ (middle), and differential in $x_{\gamma}^{\text{obs}}$ (right) as measured in \hera{2} by ZEUS~\cite{ZEUS:2007njl}. It is compared to NLO+PS predictions from the \powhegbox{} matched to parton showers. The direct and resolved contributions are shown in green and blue, respectively, and their sum in red.  Additionally the LO fixed-order sum is shown in yellow. Shaded bands indicate the envelope of a seven-point variation of the renormalization and factorization scales, while error bars indicate the numerical uncertainty.}
    \label{fig:HERA-Run-2-SaS1M}
\end{figure}

To compare with the ZEUS \hera{2} measurement~\cite{ZEUS:2007njl}, the proton beam energy is increased to \SI{920}{\giga \electronvolt} as discussed in Sec.~\ref{sec: Comparison to HERA Results}. Following the ZEUS analysis, the resolved- and direct-enriched regions are separated by $x_{\gamma}^{\text{obs}}\leq 0.75$ and $x_{\gamma}^{\text{obs}}>0.75$, respectively. We present the dijet cross sections as functions of the mean transverse energy of the two leading jets,
\begin{equation}
    \bar{E}_T=\frac{E_T^{\text{jet1}}+E_T^{\text{jet2}}}{2}~, \label{eq. ET_bar}
\end{equation}
and of the observed parton-in-photon momentum fraction $x_{\gamma}^{\text{obs}}$ as introduced in Eq.~\eqref{eq. x_gamma_obs}. In the left and middle panel of Fig.~\ref{fig:HERA-Run-2-SaS1M} we see that the central NLO+PS prediction is already compatible with data within the experimental uncertainties in almost all bins of the $\bar{E}_T$ differential cross section. Including the seven-point scale variation band, yielding an uncertainty of nearly \SI{20}{\percent}, we find good agreement between data and predictions. The predictions are below the data in the highest $\bar{E}_T$ bin in both the resolved and direct-dominated region, but these bins contain the fewest events in both the Monte Carlo event generation and the experiment, so that the two still agree within uncertainties. The LO fixed-order prediction is mostly lower by a constant factor, but drops sharply in the lowest $\bar{E}_T$ bin in both regions. This is the kinematic region where the parton shower effects are largest, since the shower predominantly adds radiation at low transverse momenta. The NLO+PS prediction correctly compensates for this dip, yielding a $K$-factor of about two relative to LO in the resolved-dominated region and a $K$-factor close to one in the direct-dominated region. Using a threshold of $x_{\gamma}^{\text{obs}}= 0.75$ proves to be an adequate choice, as in the left and middle panels of Fig.~\ref{fig:HERA-Run-2-SaS1M} the resolved and direct contributions dominate over their counterparts. This change of dominance shows even more prominently in the right panel of Fig.~\ref{fig:HERA-Run-2-SaS1M} when comparing the low and high observed parton-in-photon momentum fraction regions. Again, we notice agreement between the NLO+PS prediction and the data, in particular within the seven-point scale variation band.

\begin{figure}
    \centering
    \begin{subfigure}{0.55 \textwidth}
        \centering
        \includegraphics[width=\linewidth]{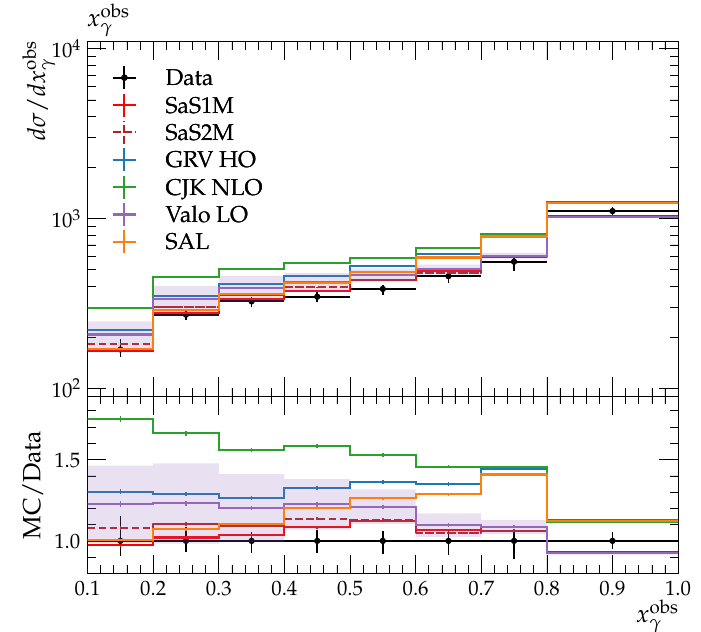}
    \end{subfigure}
    \caption{Differential dijet photoproduction cross section with respect to $x_\gamma^{\text{obs}}$ as measured in \hera{2} by ZEUS~\cite{ZEUS:2007njl}. It is compared to NLO+PS predictions from the \powhegbox{} matched to a \pythia{} parton shower. In different colors we show the impact of using different photon PDF sets. Only the \texttt{VALO} PDF set is shipped with uncertainty PDF sets, which results in a calculable PDF uncertainty, indicated by a shaded band.}
    \label{fig:HERA-Run-2-photonPDF}
\end{figure}

The distribution in $x_\gamma^{\text{obs}}$ is particularly sensitive to the photon PDF variation and is thus well-suited to compare different photon PDF sets. In Fig.~\ref{fig:HERA-Run-2-photonPDF}, the NLO+PS prediction summed over direct and resolved contributions for different photon PDFs is shown, featuring a zoomed-in ratio plot to highlight the differences. We remind the reader that an overview of the implemented photon PDF sets was given in Tab.~\ref{tab: photonPdf}. There are no error PDF sets available for any of the shown PDF sets except for \texttt{VALO10\_LO}. The corresponding uncertainty band exceeds the experimental uncertainty, in particular for small $x_\gamma^{\text{obs}}$. This is expected, since the \texttt{VALO10\_LO} is a LO fit constrained by $F_2^\gamma$ data only, which are not directly sensitive to the gluon content of the photon. Since dijet photoproduction is directly sensitive to the gluon density, including such data in future PDF fits appears promising\footnote{We note that dijet photoproduction data have recently been included in the \texttt{PL$\gamma$26} PDF set~\cite{Buckley:2026tbf}. As mentioned in Sec.~\ref{sec: framework}, this set was released only shortly before the completion of this work and is therefore not considered here.}.

Furthermore, we see the \texttt{SaS} sets, lying at the lower end of the \texttt{VALO10\_LO} uncertainty band, describe the data best, while all other setups overshoot the data, especially in the small momentum fraction region. This region is the least constrained due to high uncertainties of the gluon PDF, which is also reflected in the wider PDF uncertainty band. However, the \texttt{CJK\,NLO} prediction, exceeding the data by up to a factor of $1.75$, is clearly an outlier in this region. At increasing momentum fractions, \texttt{VALO10\_LO} comes closer to the \texttt{SaS} predictions, while \texttt{GRV\,HO} and \texttt{SAL} approach the \texttt{CJK\,NLO} line. Note that the latter three are given in the \DISg-scheme and are converted in \powhegbox, while \texttt{VALO10\_LO}, being a LO PDF set, and \texttt{SaS1M} and \texttt{SaS2M}, already given in the \MSbar-scheme, do not need to be converted. Naturally, at large  $x_{\gamma}^{\text{obs}}= 0.75$, the direct (and as such photon PDF independent) contribution becomes dominant, causing the differences between photon PDF sets to decrease. We emphasize that the deviations between different PDF sets are large and partly yield the largest contribution of the theoretical uncertainty.

\subsection{Predictions for the EIC}

The \powhegbox{} implementation presented so far is not restricted to HERA kinematics, but can also be applied directly to make predictions for the upcoming EIC. We consider two beam configurations. First, we study electron-gold collisions with beam energies of $E_e=\SI{18}{\giga \electronvolt}$ and $E_{\text{Au}}=\SI{110}{\giga \electronvolt}$ per nucleon, corresponding to a center-of-mass energy per nucleon of $\sqrt{s_{eN}}\simeq \SI{89}{\giga \electronvolt}$. This is the highest-energy $e$Au configuration considered in the EIC Yellow Report~\cite{AbdulKhalek:2021gbh}. Earlier NLO studies of dijet photoproduction in electron-nucleus collisions~\cite{Klasen:2018gtb,Guzey:2020zza} have shown that this process is sensitive to nuclear modifications of the parton densities. While keeping \texttt{SaS1M} as our default photon PDF, we compare three recent global nuclear PDF analyses, based on different fitting methodologies: \texttt{nCTEQ26}~\cite{nCTEQ:2026ewr}, building on \texttt{CJ15nlo} free-proton PDFs~\cite{Accardi:2016qay}, \texttt{EPPS21nlo\_CT18Anlo}~\cite{Eskola:2021nhw}, which parametrizes nuclear modification relative to the \texttt{CT18Anlo}~\cite{Hou:2019efy} proton baseline, and the neural-network-based \texttt{nNNPDF3.0}~\cite{AbdulKhalek:2022fyi}. All three sets include LHC proton-lead data. For gold, direct constraints in the analyses are essentially limited to fixed-target DIS conducted at JLab and SLAC and RHIC deuteron-gold data. All cross sections are given per nucleon. Second, we consider electron-proton collisions at the highest-energy scenario~\cite{AbdulKhalek:2021gbh} of $E_e=\SI{18}{\giga \electronvolt}$ and $E_p=\SI{275}{\giga \electronvolt}$, resulting in a center-of-mass energy of $\sqrt{s_{ep}}\simeq \SI{141}{\giga \electronvolt}$. This setup has been used by previous Monte Carlo generator based analyses as reported in Refs.~\cite{Meinzinger:2023xuf,Helenius:2024rth}. For this setup we compare the impact of different photon PDF sets, while keeping the proton PDF at \texttt{CJ15nlo}. In both cases, jets are reconstructed from the final state with the anti-$k_T$ jet clustering algorithm with $R=1$, demanding at least two jets with transverse momentum above $E_T>\SI{8}{\giga \electronvolt}$ in the analysis. Events are generated with a cut on the transverse momentum of the underlying born \texttt{bornktmin = 4} GeV, a factor of two below the analysis cut. As for the HERA setups, the photon virtuality in the WWA flux is restricted to a maximum of $Q^2_{\text{max}}=\SI{1}{\giga \electronvolt^2}$.

\begin{figure}
    \centering
    \begin{subfigure}{0.42 \textwidth}
        \centering
        \includegraphics[width=\linewidth]{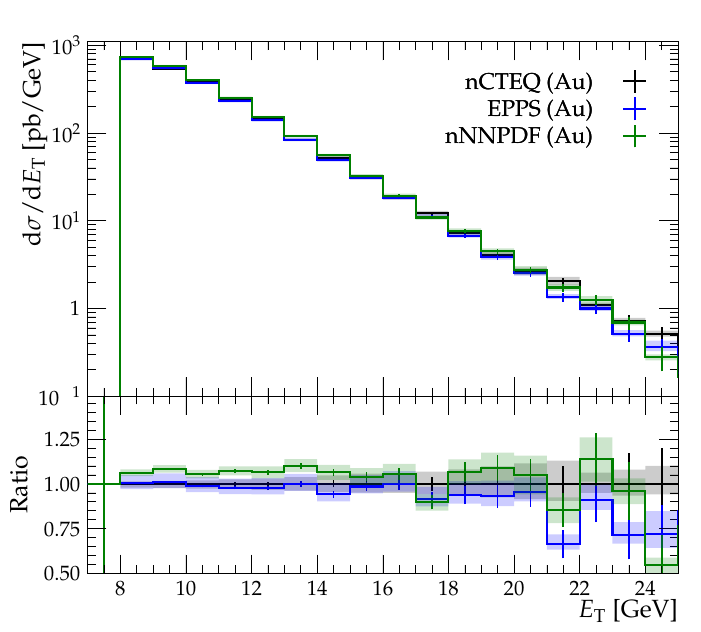}
    \end{subfigure}
    \begin{subfigure}{0.42 \textwidth}
        \centering
        \includegraphics[width=\linewidth]{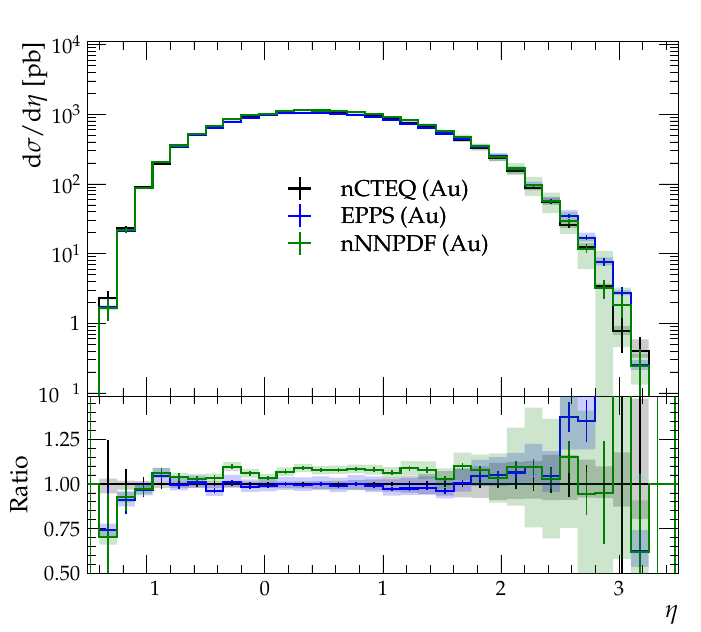}
    \end{subfigure}
    \begin{subfigure}{0.42 \textwidth}
        \centering
        \includegraphics[width=\linewidth]{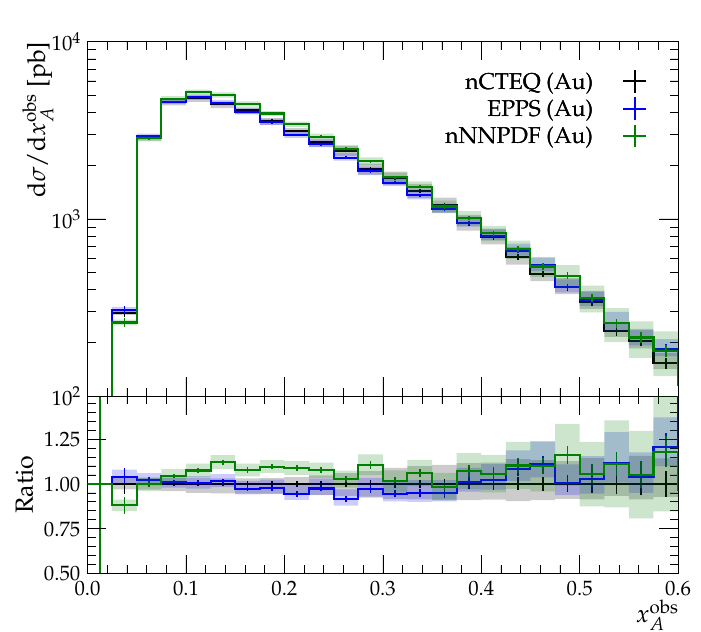}
    \end{subfigure}
    \begin{subfigure}{0.42 \textwidth}
        \centering
        \includegraphics[width=\linewidth]{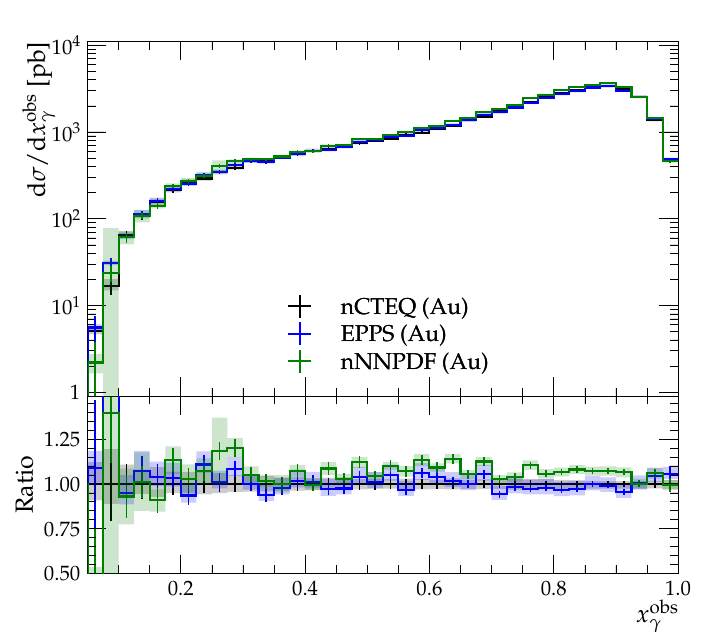}
    \end{subfigure}
    \caption{Dijet photoproduction cross section per nucleon differential in $E_T$ (top, left), $\eta$ (top, right), $x_{\text{A}}^{\text{obs}}$ (bottom, left), and $x_{\gamma}^{\text{obs}}$ (bottom, right) for electron-gold collisions as predicted at NLO+PS. We vary the nuclear PDFs between \texttt{nCTEQ26} (black), \texttt{EPPS21nlo\_CT18Anlo} (blue) and \texttt{nNNPDF3.0} (green). PDF uncertainties are shown as shaded bands and the numerical precision as error bars.}
    \label{fig: EIC nuclear}
\end{figure}

Figure~\ref{fig: EIC nuclear} shows a selection of dijet observables comparing the different nuclear PDF sets, featuring the transverse energy, the pseudorapidity and the observed momentum fractions of the photon and nucleon. The latter is defined in analogy to $x_{\gamma}^{\text{obs}}$ of Eq.~\eqref{eq:x-resolved} as
\begin{align}
    x_{\text{A}}^{\text{obs}}&= \frac{E_T^{\text{jet1}} e^{\eta^{\text{jet1}}} + E_T^{\text{jet2}} e^{\eta^{\text{jet2}}}}{2 E_\text{A}}~,
\end{align}
where $E_A$ is the beam energy per nucleon and positive rapidities point in the direction of the hadron beam. For the $ep$ setup $E_A$ and $x_{\text{A}}^{\text{obs}}$ indicate the proton energy and observed momentum fraction, respectively. With $E_T>\SI{8}{\giga \electronvolt}$ at $\sqrt{s_{eN}}\simeq \SI{89}{\giga \electronvolt}$, the accessible range is $x_{\text{A}}^{\text{obs}}\gtrsim 0.03$. In the bulk of the differential cross section, \ie~at low $E_T$ and central pseudorapidities, the \texttt{nCTEQ} and \texttt{EPPS} predictions agree within statistical uncertainty, while \texttt{nNNPDF} lies about \SIrange{5}{10}{\percent} higher, which is not always covered by the PDF uncertainties. Towards forward pseudorapidities, \ie~large momentum fractions in the nucleus, the central predictions diverge and the uncertainty bands widen substantially, most strongly for \texttt{nNNPDF}, although these regions are also limited by statistics. The same pattern appears in the $x_{\text{A}}^{\text{obs}}$ distribution. The excess of \texttt{nNNPDF} of up to \SI{10}{\percent} is concentrated at $0.1 \lesssim x_{\text{A}}^{\text{obs}} \lesssim 0.35$, \ie~the antishadowing region, whereas \texttt{EPPS} falls slightly below \texttt{nCTEQ} around $x_{\text{A}}^{\text{obs}} \approx 0.25$. The ratios between the nPDF sets are mostly in agreement within numerical and PDF uncertainties in the low $x_{\gamma}^{\text{obs}}$ region, while the differences become visible for $x_{\gamma}^{\text{obs}}\gtrsim 0.7$, approaching \SIrange{5}{10}{\percent}. Since the systematic uncertainties of EIC dijet measurements are expected at the level of a few percent~\cite{Guzey:2020zza}, such measurements could in principle discriminate between the current nPDF sets, provided the photon PDF dependence discussed below is under control.

\begin{figure}
    \centering
    \begin{subfigure}{0.42 \textwidth}
        \centering
        \includegraphics[width=\linewidth]{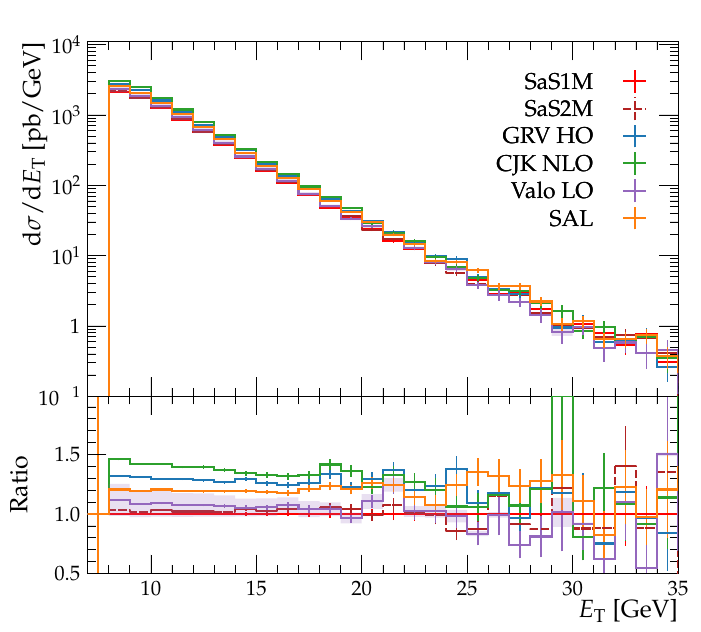}
    \end{subfigure}
    \begin{subfigure}{0.42 \textwidth}
        \centering
        \includegraphics[width=\linewidth]{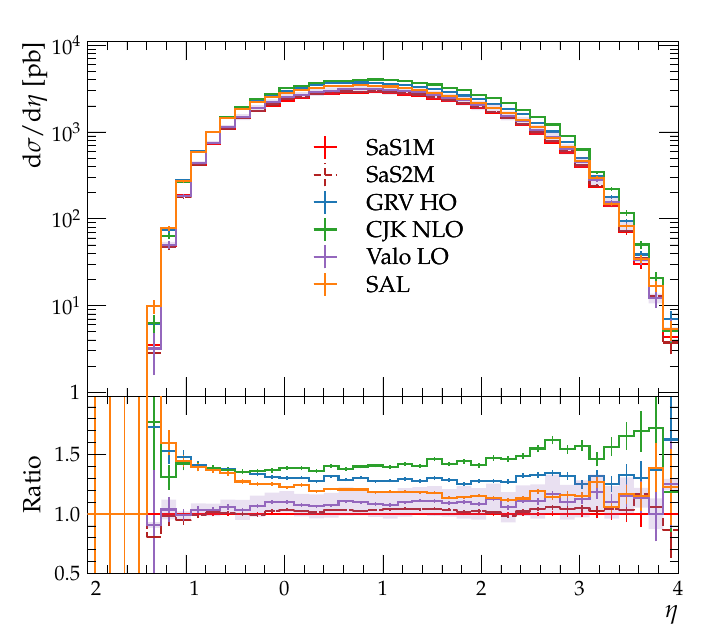}
    \end{subfigure}
    \begin{subfigure}{0.42 \textwidth}
        \centering
        \includegraphics[width=\linewidth]{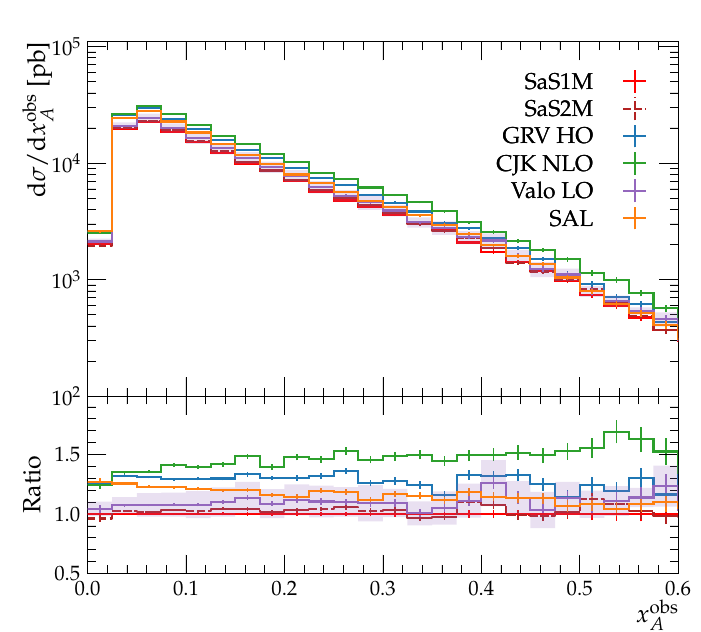}
    \end{subfigure}
    \begin{subfigure}{0.42 \textwidth}
        \centering
        \includegraphics[width=\linewidth]{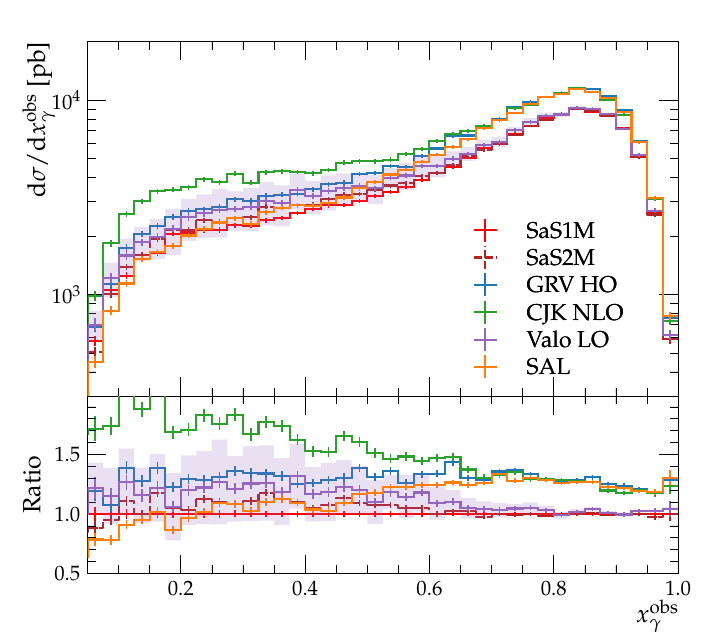}
    \end{subfigure}
    \caption{Dijet photoproduction cross section differential in $E_T$ (top, left), $\eta$ (top, right), $x_{\text{A}}^{\text{obs}}$ (bottom, left), and $x_{\gamma}^{\text{obs}}$ (bottom, right) for electron-proton collisions as predicted at NLO+PS. In different colors we show the impact of using different photon PDF sets, including their numerical errors shown as error bars. Only the \texttt{VALO} PDF set is shipped with uncertainty PDF sets, which results in a calculable PDF uncertainty, indicated by a shaded band.}
    \label{fig: EIC photon}
\end{figure}

Figure~\ref{fig: EIC photon} shows the same observables for the second setup, the proton-electron collisions with $\sqrt{s_{ep}}\simeq \SI{141}{\giga \electronvolt}$. As at \hera{2} energies, shown in Sec.~\ref{sec: hera2}, the \texttt{SaS} sets give the lowest and the \texttt{CJK\,NLO} set the highest prediction. The latter exceeds the former by about \SI{45}{\percent} at low $E_T$ and up to \SI{60}{\percent} at forward pseudorapidities. The spread decreases towards the high $E_T$ and low $\eta$ regions, which are dominated by the direct contribution. The largest differences occur at small $x_{\gamma}^{\text{obs}}$ and correspondingly at large $x_{\text{A}}^{\text{obs}}$. This is the resolved dominated region of the photon, dominated by its gluon density, which is the least constrained by data. In this region the PDF uncertainty of \texttt{VALO\_LO} is about \SI{15}{\percent}, considerably smaller than the spread to \texttt{CJK\,NLO}. At large $x_{\gamma}^{\text{obs}}$, where direct processes dominate, the sets fall into two groups, \texttt{SaS} and \texttt{VALO\_LO} on the one hand and \texttt{GRV\,HO}, \texttt{SAL}, and \texttt{CJK\,NLO} on the other, separated by about \SI{25}{\percent}. Given the large direct contribution, this reflects sizable differences in the parton densities at large $x$.

Comparing Figs.~\ref{fig: EIC nuclear} and~\ref{fig: EIC photon}, and keeping in mind the different beam energies, the photon PDFs induce considerably larger uncertainties than the differences between nPDF sets. This analysis therefore implies, that predictions for dijet photoproduction in $e$Au are currently limited by the photon PDFs. A better determination of the photon PDFs, e.g.~from HERA and EIC $ep$ photoproduction data, is needed for photoproduction to contribute to global nPDF fits. Whether the smaller uncertainty of \texttt{VALO\_LO} reflects a genuine improvement remains to be tested against data, and the NLO+PS implementation in \powhegbox{} presented here provides a suitable tool for such comparisons.

\section{Conclusion}
\label{sec: conclusion}

In this paper, we presented the new implementation of dijet photoproduction in electron-proton and electron-nucleus scattering within the \powhegbox{} framework. Our implementation includes a dedicated phase space parametrization that handles both direct and resolved photon contributions, the corresponding analytic matrix elements, and a consistent treatment of the separation of these two components in NLO of perturbative QCD.

Our implementation was validated with several complementary checks. First, we demonstrated that the dependence of the cross section on the photon factorization scale $\mu_F^{(\gamma)}$ is largely canceled by the collinear $\gamma \rightarrow q \bar{q}$ splitting contribution, as expected from factorization theorems. Second, we verified that the treatment of asymmetric beam configurations is consistent by imposing symmetry under beam exchange for observables that are invariant under this exchange. Finally, we compared our fixed-order results with \sherpa{} and found agreement within numerical uncertainties for two-jet inclusive observables as well as for three-jet observables, which test exclusively the real-emission phase space. 

At NLO matched to parton showers, we studied the impact of the different shower components on the dijet cross section. The migration over the fiducial cuts increases the cross section by about \SI{30}{\percent} with ISR being the most prominent contribution, responsible for an increase of about \SI{50}{\percent} in resolved-enriched regions. We also investigated different parton-shower matching setups by varying the matching scale and the hardness definition via the \hooks{} in \pythia. The observed impact of up to \SI{30}{\percent} should be kept in mind when interpreting further results, as it constitutes a source of theoretical uncertainty.

Comparing the showered predictions with data from \hera{1} and \hera{2}, we found agreement within the theoretical uncertainty obtained from a seven-point scale variation. However, the central prediction tends to slightly overshoot the data in the resolved-dominated region and undershoot it in the direct-dominated region. Moreover, the comparison between different photon PDFs revealed discrepancies of up to \SI{75}{\percent} in the low-$x$ region, motivating a new examination of the photon structure that follows up on the efforts of, e.g., the new \texttt{VALO} analysis before the EIC starts running.

Finally, we presented predictions for electron-gold and electron-proton collisions at beam energies expected for the EIC. We find that the \texttt{nNNPDF3.0} set globally exceeds the predictions of \texttt{nCTEQ26} and \texttt{EPPS21} by about \SIrange{5}{10}{\percent}. The variation between photon PDF sets in $ep$ collisions shows differences of up to \SI{60}{\percent} in the $E_T$ and $\eta$ spectrum and even larger in the low-$x_\gamma^{\text{obs}}$ region, with \texttt{SaS} usually being the lowest and \texttt{CJK\,NLO} yielding the highest prediction.

In summary, this paper demonstrates the successful implementation of jet photoproduction in \powhegbox{}, which therefore now provides a flexible tool for the interpretation of present and future photoproduction data. This code is publicly available as the \texttt{photoproduction-dijet} submodule of the \powhegbox{} \texttt{User-Processes} repository on GitLab.

\appendix
\section{\pythia{} settings}
\label{sec: appendix Pythia8 settings}

In this appendix, we list exemplary settings for \pythia{} for an NLO $ep$ run with the PDFs of the proton set to \texttt{CJ15nlo}. Some settings that are kept to their default value, but are of importance, are listed for completeness. To distinguish them, they are grayed out. Comments are marked with a \texttt{!}.

\begin{Verbatim}[commandchars=\\\{\}]
    Beams:frameType = 4
    PDF:beamA2gamma = on
    PDF:lepton2gamma = on
    PDF:pSet = LHAPDF6:CJ15nlo

    \textcolor{gray}{Photon:ProcessType = 0}
    \textcolor{gray}{PDF:lepton2gammaSet = 2}
    \textcolor{gray}{Photon:sampleQ2 = on}
    \textcolor{gray}{Photon:Q2max = 1.0}
    Photon:Wmin = 5                     !minimum value

    \textcolor{gray}{PartonLevel:ISR = on}
    \textcolor{gray}{PartonLevel:FSR = on}
    PartonLevel:MPI = off
    
    SpaceShower:QEDshowerByQ = off      !on/off for resolved/direct
    SpaceShower:QEDshowerByL = off
    TimeShower:QEDshowerByQ = off   
    TimeShower:QEDshowerByL = off   

    TimeShower:pTmaxMatch = 2
    SpaceShower:pTmaxMatch = 2

    ! POWHEG Hooks
    POWHEG:nFinal = 2
    POWHEG:veto = 1
    POWHEG:vetoCount = 50
    \textcolor{gray}{POWHEG:pThard = 0}
    POWHEG:pTdef = 1
    \textcolor{gray}{POWHEG:pTemt = 0}
    \textcolor{gray}{POWHEG:emitted = 0}
    POWHEG:QEDveto = 1
\end{Verbatim}

\acknowledgments
The authors thank Peter Meinzinger for his help in setting up and validating the \sherpa{} predictions and Ilkka Helenius for his support in adapting the \pythia{} shower to handle the photoproduction events generated with \powhegbox{}. Calculations for this publication were performed on the HPC cluster PALMA~\Romannum{2} of the University of Münster, subsidized by the DFG (INST 211/667-1).

\bibliographystyle{JHEP}
\bibliography{References.bib}

\end{document}